\documentclass[journal=jctcce,manuscript=article,layout=twocolumn]{achemso}

\usepackage[colorlinks,linkcolor=blue,citecolor=blue,urlcolor=black,bookmarks=false,hypertexnames=true]{hyperref} 

\usepackage[version=3]{mhchem}
\usepackage{amssymb}
\usepackage{amsmath}
\usepackage[utf8]{inputenc}
\usepackage{textcomp}
\usepackage{color}
\usepackage{braket}
\usepackage{xspace}
\usepackage[normalem]{ulem}

\usepackage{graphicx}
\usepackage{subfigure}
\usepackage{threeparttable}
\usepackage{textcomp}
\usepackage{setspace}
\usepackage{caption}
\usepackage{comment}
\usepackage{multirow}
\usepackage{cleveref}
\usepackage{booktabs}

\usepackage{array}
\newcolumntype{L}[1]{>{\raggedright\let\newline\\\arraybackslash\hspace{0pt}}m{#1}}
\newcolumntype{C}[1]{>{\centering\let\newline\\\arraybackslash\hspace{0pt}}m{#1}}
\newcolumntype{R}[1]{>{\raggedleft\let\newline\\\arraybackslash\hspace{0pt}}m{#1}}

\makeatletter
\let\l@addto@macro\relax
\makeatother
\usepackage[fontsize=11pt]{scrextend}

\let\oldmaketitle\maketitle
\let\maketitle\relax

\newcommand{\cm}{\ensuremath{\text{cm}^{-1}}\xspace}

\crefname{figure}{Figure}{Figures}
\crefname{table}{Table}{Tables}
\crefname{equation}{Eq.}{Eqs.}
\crefname{section}{Section}{Sections}
\crefname{subsection}{Section}{Sections}

\author{Nicholas~Yiching~Chiang}
\affiliation{Department of Chemistry and Biochemistry, The Ohio State University, Columbus, Ohio 43210, USA}

\author{Rajat~Majumder}
\affiliation{Department of Chemistry, University of Washington, Seattle, Washington 98195, USA}

\author{Terrence~L.~Stahl}
\affiliation{Department of Chemistry, Brown University, Providence, Rhode Island 02912, USA}

\author{Ning-Yuan~Chen}
\affiliation{Department of Chemistry, Yale University, New Haven, Connecticut 06511, USA}

\author{Alexander~Yu.~Sokolov}
\email{sokolov.8@osu.edu}
\affiliation{Department of Chemistry and Biochemistry, The Ohio State University, Columbus, Ohio 43210, USA}

\begin{tocentry}
\includegraphics[width=1.0\textwidth]{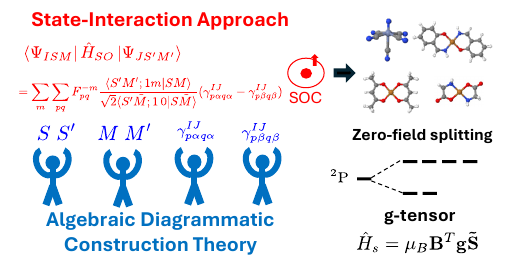}
\end{tocentry}

\title{{\color{blue}
Efficient Spin–Orbit Coupling in \\Algebraic Diagrammatic Construction Theory: \\A State--Interaction Approach
}}
\begin{document}

\newcommand*{\abstractext}{
Spin--orbit coupling and electron correlation play important roles in a broad range of chemical and physical phenomena. In this work, we systematically assess the performance of single-reference algebraic diagrammatic construction combined with state interaction (SI-ADC) for describing electronic structure in the presence of spin--orbit coupling. We find that second-order SI-ADC methods (SI-ADC(2)) achieve zero-field-splitting accuracies comparable to those of higher-level and more computationally demanding approaches, including variational four-component ADC and two-component ADC methods based on spin--orbit effective Hamiltonians. We further demonstrate the broad applicability of SI-ADC by evaluating a diverse range of spin--orbit-coupled properties, including electron affinities, ionization potentials, neutral excitation energies, core-ionization energies, and magnetic $g$-tensors, for systems ranging from small molecules to transition-metal complexes and $f$-block compounds. Across these applications, SI-ADC(2) methods provide consistently accurate results, likely benefiting in part from favorable error cancellation between electron-correlation and spin--orbit effects. In contrast, SI-ADC(3) methods are generally less accurate, particularly for systems containing transition-metal and heavy elements. Overall, these results establish SI-ADC as an efficient and broadly applicable framework for incorporating spin--orbit coupling into ADC calculations of electronic and spectroscopic properties.
\vspace{0.25cm}
}

\twocolumn[
\begin{@twocolumnfalse}
\oldmaketitle
\vspace{-0.75cm}
\begin{abstract}
\abstractext
\end{abstract}
\end{@twocolumnfalse}
]


\section{Introduction}
\label{sec:introduction}
Spin–orbit coupling (SOC), arising from relativistic effects\cite{truhlar:2025p4301,pyykko:2012p45,saue:2011p3077,khan:2015p,maurice:2017p765}, plays a fundamental role in a wide range of chemical and physical phenomena, including zero-field splitting (ZFS)\cite{wu:2022p2199}, magnetic properties\cite{atanasov:2015p177,staab:2022p6588,omist:2025p1153,perfetti:2019p11875,maurice:2013p18784,campanella:2024p11}, intersystem crossing,\cite{pope:2024p1337,penfold:2018p6975,marian:2021p617,marian:2012p187,kim:2020p621} and X-ray fine structure\cite{mandal:2026p044105,vidal:2020p8314,kasper:2020p,wasinger:2003p12894}. Reliable theoretical descriptions of SOC provide valuable mechanistic insights into these properties and are essential for understanding and predicting their behavior. Such capabilities have enabled advances in diverse applications, ranging from the design of efficient photoredox catalysts\cite{koike:2023p100205,poh:2025p36579,dong:2025p17276} to the development of single-molecule magnets (SMMs)\cite{zadrozny:2013p125,zabala-lekuona:2021p213984,suturina:2015p9948,reta:2021p5943} for next-generation information storage. Despite its importance, a reliable theoretical treatment of SOC remains a significant challenge because it requires both an accurate description of relativistic effects and a balanced treatment of electron correlation in multielectron systems.

To address these challenges, two general strategies have been developed for incorporating SOC into electronic structure calculations: variational and perturbative approaches. In variational methods\cite{ganyushin:2013p104113,sun:2020p4533,tang:2024p9917,wu:2025p244114,hess:1986p3742,jorgenaa.jensen:1996p4083,reynolds:2019p1560,wang:2026p3988,liu:2024p084111,yang:2026p014105,banerjee:2026p016101}, a four-component\cite{jorgenaa.jensen:1996p4083,thyssen:2008p034109} or two-component Dirac Hamiltonian \cite{zhang:2024p3408,bruder:2023p194117} is employed during the variational calculation, providing a rigorous treatment of relativistic and spin--orbit effects. However, when combined with high-level electron correlation methods, the computational cost of variational SOC treatments increases rapidly, limiting their applicability as molecular size grows. Moreover, because relativistic effects are incorporated directly into the electronic Hamiltonian, variational treatments require modifications to the underlying electronic structure formalism. Consequently, each electronic structure method generally requires a model-specific implementation, limiting the transferability of the approach across different computational frameworks. 

On the other hand, the perturbative approach\cite{majumder:2024p4676,majumder:2023p546,cheng:2014p164107} introduces the SOC effect as a perturbation on top of a spin-free calculation. This strategy effectively decouples the treatment of SOC corrections from the initial electronic structure evaluation, offering reasonable computational cost.  One of the most widely used perturbative methods for treating SOC is the state--interaction (SI) approach\cite{carreras:2020p214107,cebreiro-gallardo:2025p6528,jangid:2026p318,malmqvist:2002p230,atanasov:2015p177,ganyushin:2006p024103,liao:2023p358}. Based on quasidegenerate perturbation theory\cite{hose:1982p2133,lindgren:1974p2441}, the SI approach constructs a perturbation spin–orbit Hamiltonian in the basis of spin-free eigenstates. Diagonalization of this effective Hamiltonian then yields the SOC energies and wavefunctions. Owing to its flexibility, the SI approach has been successfully integrated into a variety of electronic structure methods, including multireference configuration interaction (MRCI)\cite{ganyushin:2008p114117,xiao:2024p1776}, complete active space second-order perturbation theory (CASPT2)\cite{bolvin:2006p1575,vancoillie:2007p1803,havlas:1999p2299}, N-electron valence perturbation theory (NEVPT2)\cite{majumder:2023p546,singh:2018p4662,sharma:2017p6906,lang:2020p014109,nehrkorn:2018p15330,atanasov:2011p7460}, 
linearized pair-density functional theory (L-PDFT)\cite{jangid:2026p318}, 
second-order multireference driven similarity renormalization group perturbation theory (DSRG-MRPT2)\cite{wang:2025p2082,park:2025p12071}, 
iterative configuration interaction (iCI)\cite{zhang:2022p224007},
and equation-of-motion coupled-cluster (EOM-CC)\cite{pokhilko:2019p034106,alessio:2021p4225,alessio:2023p3647}. Nevertheless, because the SI approach treats SOC perturbatively, its accuracy may deteriorate for systems with strong spin–orbit coupling, where the perturbative treatment of SOC is no longer adequate.

In addition to accurately describing SOC, practical quantum chemical methods must provide a balanced treatment of electron correlation across many electronic states.
The algebraic diagrammatic construction (ADC) \cite{schirmer:1982p2395,schirmer:1983p1237,schirmer:1991p4647,mertins:1996p2140,trofimov:2006p1,dreuw:2015p82,banerjee:2023p3037,dreuw:2023p119} theory provides an attractive platform for achieving this goal.
ADC offers a systematically improvable hierarchy of perturbative approximations that efficiently incorporates electron correlation while providing access to large manifolds of electronic states.
Moreover, different formulations of ADC provide a unified framework for describing neutral and charged excitations as well as valence and core-level spectroscopies, making ADC particularly well suited for the development of broadly applicable approaches to relativistic electronic structure.
Over the past decades, ADC methods have been successfully applied to a wide range of spectroscopic properties, including UV/Vis\cite{stahl:2022p044106,papapostolou:2026p114101,papapostolou:2025p234123,brand:2024p103}, two-photon circular dichroism (TPCD)\cite{papapostolou:2025p6133}, magnetic properties\cite{schneider:2023p8723,fedotov:2022p174109}, X-ray spectra,\cite{mazin:2023p4991,ahmed:2025p7588,demoura:2024p5816,kaczun:2023p5648,dreuw:2022p11259,brumboiu:2021p044106,fransson:2019p546,rehn:2017p5552,wenzel:2015p214104} charged excitations,\cite{stahl:2024p204104,banerjee:2021p074105,banerjee:2023p3037,demoura:2024p5816,banerjee:2019p224112,muller:2026p181102,weidlich:2025pe70095,rehn:2024p8795,leitner:2024p7680,dreuw:2023p6635} and electronic processes in condensed phase environments.\cite{serna:2025p1156,scheurer:2018p4870,scheurer:2019p6154,scheurer:2021p3445,marefatkhah:2018p4640,prager:2016p204103,banerjee:2022p5337,ahmed:2025p7588}

Recent developments have extended the ADC framework to incorporate SOC effects.
In particular, variational treatments of SOC have been implemented within ADC,\cite{chakraborty:2026p3971,chakraborty:2025p104106,pernpointner:2018p1510,pernpointner:2014p084108,nikoobakht:2015p3431,brandt:2015p7,pernpointner:2010p205102} providing an accurate description of spin--orbit interactions.
To develop a more computationally efficient alternative, our group recently introduced a complementary perturbative approach\cite{majumder:2025p2414} that incorporates SOC directly into the ADC effective Hamiltonian, enabling relativistic effects and dynamic electron correlation to be treated simultaneously on equal footing.
An alternative strategy is provided by the state--interaction (SI) formulation of ADC, which has demonstrated promising performance for describing spin--orbit couplings between singlet and triplet states.\cite{krauter:2017p286}
Compared with the variational and effective-Hamiltonian formulations, however, the capabilities and limitations of SI-ADC have received substantially less systematic investigation.
In particular, its accuracy across different ADC truncation levels, classes of electronic states, and regimes ranging from weak to strong SOC has not yet been comprehensively assessed.

In this work, we systematically investigate the capabilities of the perturbative SI approach for describing SOC effects within the ADC framework.
We benchmark SI-ADC against complementary ADC approaches employing different treatments of SOC to assess its accuracy and numerical stability and to establish the range of problems for which it provides a computationally efficient alternative.
To demonstrate the generality of SI-ADC approach, we consider a broad spectrum of electronic and spectroscopic properties, including electron affinities, ionization potentials, neutral excitation energies, core ionization energies, zero-field splittings, and magnetic $g$-tensors.
Our benchmark set spans systems ranging from small main-group molecules to transition-metal complexes and molecules containing heavy $f$-block elements, probing SI-ADC across widely varying strengths of electron correlation and SOC.
Through these comparisons, we assess the performance and practical advantages of different SI-ADC approximations and examine how their accuracy evolves with perturbation order and increasing relativistic effects.

\section{Theory}
\label{sec:theory}

\subsection{Algebraic Diagrammatic Construction Theory}
\label{sec:theory:ADC}
 
We begin with a brief overview of algebraic diagrammatic construction (ADC) theory,\cite{schirmer:1982p2395,schirmer:1983p1237,schirmer:1991p4647,mertins:1996p2140,trofimov:2006p1,dreuw:2015p82,banerjee:2023p3037} which provides a hierarchy of approximations to a many-body propagator. In its general frequency-dependent form, the propagator can be written as:\cite{danovich:2011p377,dickhoff:2005p,fetter:2003p,schirmer:2018p}
\begin{align}
  \label{eq:1gf}
G_{\mu \nu}(\omega) &= G^{+}_{ \mu \nu}(\omega) \pm G^{-}_{\mu \nu}(\omega) \notag \\
&= \langle{\Psi^{N}_{0}}|q_{\mu } (\omega - \hat{H} + E^{N}_{0})^{-1} q^{\dagger}_{\nu} |{\Psi^{N}_{0}}\rangle \notag \\
&+ \langle{\Psi^{N}_{0}}|q^{\dagger}_{\nu} (\omega + \hat{H} - E^{N}_{0})^{-1} q_{\mu} |{\Psi^{N}_{0}}\rangle
\end{align}
In \cref{eq:1gf}, $G^{+}_{ \mu \nu}(\omega)$ and $G^{-}_{ \mu \nu}(\omega)$ denote the forward and backward components of the propagator, respectively, $\hat{H}$ is the electronic Hamiltonian, $\ket{\Psi^{N}_{0}}$ and $E^{N}_{0}$ are the $N$-electron reference state and its energy, and $\omega$ represents the frequency of external radiation. The operators $q^{\dagger}_{\nu}$ and $q_{\mu }$ denote the perturbation and observable operators, respectively. Different choices of $q_{\mu}$ and $q^{\dagger}_{\nu}$ enable the description of different spectroscopic processes within the same general propagator framework.

In this work, we consider three types of excitation processes. The first two correspond to charged excitations: electron attachment (EA) and ionization (IP). For these processes, the operators are defined as $q_{\mu}=a_p$ and $q_{\nu}^{\dagger}=a_q^{\dagger}$. With this choice, \cref{eq:1gf} can be expressed as:
\begin{align}
\label{eq:spectral_gf_IPEA}
G_{pq}^{\mathrm{EA/IP}}(\omega) &= \sum_n \frac{\langle{\Psi^{N}_{0}}|a_p|{\Psi^{N+1}_n}\rangle \langle{\Psi^{N+1}_n}|a^{\dagger}_q|{\Psi^{N}_{0}}\rangle}{\omega - E^{N+1}_{n} + E^{N}_{0}} \notag \\
			&+ \sum_n \frac{\langle{\Psi^{N}_{0}}|a^{\dagger}_q|{\Psi^{N-1}_n}\rangle \langle{\Psi^{N-1}_n}|a_p|{\Psi^{N}_{0}}\rangle}{\omega + E^{N-1}_{n} - E^{N}_{0}} 
\end{align}
In \cref{eq:spectral_gf_IPEA}, the first term corresponds to electron attachment, whereas the second describes electron detachment (ionization). 

The third type corresponds to neutral electronic excitations (EE). These can be described by introducing fluctuation operators $q_{\mu} = a^{\dagger}_{q}a_{p} - \langle \Psi^{N}_{0} | a^{\dagger}_{q}a_{p} | \Psi^{N}_{0} \rangle$ and $q_{\nu}^{\dagger} = a^{\dagger}_{r}a_{s} - \langle \Psi^{N}_{0} | a^{\dagger}_{r}a_{s} | \Psi^{N}_{0} \rangle$, leading to the polarization propagator:
\begin{align}
\label{eq:spectral_gf_EE}
G_{pq,rs}^{\mathrm{EE}}(\omega) \equiv \Pi_{pq, rs}(\omega) &= \sum_{n>0} \frac{\langle{\Psi^{N}_{0}}| a^{\dagger}_{q}a_p |{\Psi^{N}_n}\rangle \langle{\Psi^{N}_n}| a^{\dagger}_r a_{s} |{\Psi^{N}_{0}}\rangle}{\omega - E^{N}_{n} + E^{N}_{0}} \notag \\
			&+ \sum_{n>0} \frac{\langle{\Psi^{N}_{0}}| a^{\dagger}_r a_{s} |{\Psi^{N}_n}\rangle \langle{\Psi^{N}_n}| a^{\dagger}_{q}a_p |{\Psi^{N}_{0}}\rangle}{\omega + E^{N}_{n} - E^{N}_{0}} 
\end{align}
Since the backward and forward components of the polarization propagator are related via $\boldsymbol{\Pi}^{-}(\omega) = (\boldsymbol{\Pi}^{+}(-\omega))^{\dagger}$, only the forward component needs to be approximated in EE-ADC.

In EA-, IP-, and EE-ADC calculations, the propagator contributions in \cref{eq:spectral_gf_IPEA,eq:spectral_gf_EE} are expressed in matrix form as
\begin{align}
  \label{eq:1gf_matrix}
\mathbf{G}_{\pm}(\omega) = \mathbf{T}_{\pm} (\omega \mathbf{1}-\mathbf{M}_{\pm})^{-1}\mathbf{T}^{\dagger}_{\pm}
\end{align}
where $\mathbf{M}_{\pm}$ denotes the effective Hamiltonian and $\mathbf{T}_{\pm}$ represents the effective transition moment matrix. 
The $\mathbf{M}_{\pm}$ and $\mathbf{T}_{\pm}$ matrices can be expanded according to the Møller--Plesset perturbation theory\cite{moller:1934p618} series through order $n$,
\begin{align}
	\label{eq:M_series}
		&\mathbf{M}_{\pm} \approx  \mathbf{M}^{(0)}_{\pm} + \mathbf{M}^{(1)}_{\pm} + \mathbf{M}^{(2)}_{\pm}  \cdots+ \mathbf{M}^{(n)}_{\pm} 
		\\ 
		&\mathbf{T}_{\pm} \approx  \mathbf{T}^{(0)}_{\pm} + \mathbf{T}^{(1)}_{\pm} + \mathbf{T}^{(2)}_{\pm} \cdots + \mathbf{T}^{(n)}_{\pm}
	\label{eq:T_series}
\end{align}
yielding the $n$th-order ADC method (ADC($n$)). 

Diagonalization of the $\mathbf{M}_{\pm}$ matrix yields the eigenvalue matrix $\mathbf{\Omega}_{\pm}$, whose elements correspond to the excitation energies: 
\begin{align} 
	\label{eq:M_eig} 
	\mathbf{M}_{\pm}\mathbf{Y}_{\pm} = \mathbf{Y}_{\pm} \mathbf{\Omega}_{\pm} 
\end{align} 
The resulting eigenvector matrix $\mathbf{Y}_{\pm}$ can be used to compute the approximate spectroscopic amplitudes via $\mathbf{X}_{\pm}=\mathbf{T}_{\pm}\mathbf{Y}_{\pm}$, from which transition intensities and other spectroscopic properties can be obtained.

\subsection{State--Interaction Treatment of Relativistic Effects}
\label{sec:theory:SOC}

To incorporate relativistic effects within the ADC framework, we introduce the two-component relativistic Hamiltonian\cite{barysz:2002p2696,dyall:2020p}:
\begin{align}
	\label{eq:H_2c}
	\hat{H}_{\mathrm{2c}} = \hat{H}_{\mathrm{SF}} + \hat{H}_{\mathrm{SO}}
\end{align}
where $\hat{H}_{\mathrm{SF}}$ represents the scalar relativistic Hamiltonian, while $\hat{H}_{\mathrm{SO}}$ accounts for spin--orbit coupling. Here, the exact two-component spin-free one-electron Hamiltonian\cite{liu:2009p031104,li:2012p154114}, $\hat{H}_{\mathrm{SF}}^{\mathrm{X2C\text{-}1e}}$, is adopted as $\hat{H}_{\mathrm{SF}}$. Scalar relativistic effects are therefore incorporated directly into the reference self-consistent field (SCF) calculation.

For the spin-dependent Hamiltonian $\hat{H}_{\mathrm{SO}}$, we consider two forms: the Breit--Pauli (BP) Hamiltonian\cite{breit:1932p616,bethe:1957p} and the exact two-component first-order Douglas--Kroll--Hess approach (sf-X2C-1e + so-DKH1),\cite{liu:2009p031104,li:2012p154114} which we refer to as the XDKH1 approach. The Breit--Pauli Hamiltonian is attractive because of its compact form and relatively low computational cost. However, its accuracy deteriorates for systems containing fourth-row and heavier elements.\cite{majumder:2023p546,majumder:2024p4676} 
In contrast, the XDKH1 approach employed in this work is based on the spin-free exact two-component framework developed by Liu and co-workers (X2C-1e)\cite{liu:2009p031104,li:2012p154114} and provides a more accurate treatment of relativistic effects in molecules containing heavier elements.

To incorporate spin--orbit coupling (SOC) into the spin-free electronic structure framework, we employ the state--interaction algorithm implemented by Pokhilko and co-workers.\cite{pokhilko:2019p034106, carreras:2020p214107} In this approach, the spin--orbit Hamiltonian is introduced as a first-order perturbation in the basis of spin--free eigenstates. The matrix elements of the total electronic Hamiltonian, including the SOC contribution, are therefore expressed as:
\begin{align}
	\label{eq:SI_SOC}
	H_{ISM,JS'M'} = E_{I}\delta_{IJ}\delta_{SS'}\delta_{MM'}+\bra{\Psi_{ISM}}\hat{H}_{SO}\ket{\Psi_{JS'M'}}
\end{align}
where $\ket{\Psi_{ISM}}$ denotes the $I$th spin-free eigenstate with spin quantum number $S$ and spin projection $M$, and $E_I$  is its corresponding energy. The operator $\hat{H}_{\mathrm{SO}}$ represents the spin--orbit coupling Hamiltonian in \cref{eq:H_2c}. Diagonalization of \cref{eq:SI_SOC} yields the spin--orbit-coupled energies and wavefunctions.

In this work, $\hat{H}_{\mathrm{SO}}$ is approximated using the spin--orbit mean-field (SOMF) approximation,\cite{hess:1996p365,berning:2000p1823} giving:
\begin{align}
	\label{eq:SOMF}
	\hat{H}_{\mathrm{SO}} = \sum_{m}\sum_{pq} F^{-m}_{pq}\hat{T}^{1,m}_{pq}
\end{align}
Here, $m = -1, 0, 1$ labels the spherical tensor components, and $(p, q, r, s)$ denote spatial molecular orbitals. The quantities $F^{-m}_{pq}$ are the spherical components of the spin--orbit Hamiltonian obtained from the BP or XDKH1 Hamiltonian, while $\hat{T}^{1,m}_{pq}$ are the corresponding rank-one spherical tensor operators:
\begin{align}
	&\hat{T}^{1,-1}_{pq} = a^{\dagger}_{p\beta}a_{q\alpha}
	\label{eq:triplet_tensor_m1}
	\\ 
	&\hat{T}^{1,0}_{pq} = \frac{1}{\sqrt{2}}(a^{\dagger}_{p\alpha}a_{q\alpha} - a^{\dagger}_{p\beta}a_{q\beta})
	\label{eq:triplet_tensor_0}
	\\ 
	&\hat{T}^{1,1}_{pq} = -a^{\dagger}_{p\alpha}a_{q\beta}
	\label{eq:triplet_tensor_p1}
\end{align}

To efficiently evaluate the SOC matrix elements in \cref{eq:SI_SOC}, we apply the Wigner--Eckart theorem.\cite{wigner:1927p624,eckart:1930p305} Matrix elements involving the spherical spin excitation operators can then be expressed as:
\begin{align}
	\label{eq:Wigner}
	\bra{\Psi_{ISM}}\hat{T}^{1,m}_{pq}\ket{\Psi_{JS'M'}} = \langle S'M';1m|SM \rangle \braket{IS||\hat{T}^{(1)}_{pq}||JS'} 
\end{align}
where $\braket{S'M'; 1m|SM}$ denotes the Clebsch--Gordan (CG) coefficient~\cite{alex:2011p023507}, and $\braket{IS||\hat{T}^{(1)}_{pq}||JS'}$ is the corresponding one-particle reduced matrix element (1-RDM), which is independent of the spin-projection quantum numbers $M$, $M'$, and $m$. In this work, this quantity is referred to as the ``Wigner 1-RDM''. As shown in \cref{eq:Wigner}, matrix elements for different values of $M$, $M'$, and $m$ can therefore be reconstructed from the same Wigner 1-RDM using the appropriate CG coefficients. Substituting \cref{eq:Wigner} into the second term of \cref{eq:SI_SOC} gives:
\begin{align}
	\label{eq:soc_matrix}
	&\bra{\Psi_{ISM}}\hat{H}_{SO}\ket{\Psi_{JS'M'}} \notag \\
	&= \sum_{m}  \langle S'M';1m|SM \rangle \sum_{pq}    F^{-m}_{pq} \braket{IS||\hat{T}^{(1)}_{pq}||JS'}
\end{align}

To obtain the Wigner 1-RDM, we use spin-projection states satisfying $M = M' \equiv \tilde{M}$ and select the spherical component $m = 0$. The Wigner 1-RDM can then be expressed as:
\begin{align}
	\label{eq:Wigner_rdm}
	\braket{IS||\hat{T}^{(1)}_{pq}||JS'}  = \frac{\gamma^{IJ}_{p\alpha q\alpha} - \gamma^{IJ}_{p \beta  q \beta}}{\sqrt{2} \langle S'\tilde{M};1\,0 |S\tilde{M} \rangle}
\end{align}
Here, $\gamma^{IJ}_{p\alpha q\alpha}=\braket{\Psi_{IS\tilde{M}}|a^\dag_{p\alpha}a_{q\alpha}|\Psi_{JS'\tilde{M}}}$ and $\gamma^{IJ}_{p \beta  q \beta} = \braket{\Psi_{IS\tilde{M}}|a^\dag_{p\beta}a_{q\beta}|\Psi_{JS'\tilde{M}}}$ represent the $\alpha$- and $\beta$-spin sectors of the one-particle transition density matrices, respectively. Combining \cref{eq:Wigner_rdm,eq:soc_matrix} yields the final working expression for the SOC matrix elements:
\begin{align}
	\label{eq:soc_working_equation}
	&\bra{\Psi_{ISM}}\hat{H}_{SO}\ket{\Psi_{JS'M'}}  \notag \\
	&= \sum_{m} \sum_{pq} F^{-m}_{pq} \frac{\langle S'M';1m|SM \rangle }{\sqrt{2} \langle S'\tilde{M};1\,0 |S\tilde{M} \rangle} ( \gamma^{IJ}_{p\alpha q\alpha} - \gamma^{IJ}_{p \beta  q \beta})
\end{align}

\cref{eq:soc_working_equation} shows that SOC matrix elements between states $I$ and $J$ with arbitrary spin projections can be evaluated from transition 1-RDMs constructed between states with a common spin projection $\tilde{M}$. The dependence on $M$, $M'$, and $m$ is then recovered entirely through the corresponding CG coefficients. This formulation avoids the explicit evaluation of transition density matrices for every combination of spin projections and provides an efficient route for constructing the complete state--interaction SOC Hamiltonian.

\subsection{State--Interaction ADC}
\label{sec:theory:SIADC}

Application of the SI approach within the ADC framework requires, for each pair of interacting states, the spin quantum numbers ($S$, $S'$, $\tilde{M}$) and the spin-resolved transition one-particle reduced density matrices (1-RDMs), $\gamma^{IJ}_{\alpha\alpha}$ and $\gamma^{IJ}_{\beta\beta}$. These quantities provide the information needed to construct the Wigner 1-RDMs and evaluate the SOC matrix elements in \cref{eq:soc_working_equation}.

The spin-resolved transition 1-RDMs were derived and implemented in the ADC module of the PySCF package\cite{sun:2026p102502} for calculations based on unrestricted Hartree--Fock (UHF) reference wavefunctions, following the approach described in Refs.\@ \citenum{stahl:2022p044106}, \citenum{stahl:2024p204104} and \citenum{stahl:2026p209901}. 
For neutral excitations within EE-ADC, our implementation additionally supports the evaluation of $\gamma^{IJ}_{\alpha\alpha}$ and $\gamma^{IJ}_{\beta\beta}$ using a ROHF reference wavefunction.

For a UHF reference, $\gamma^{IJ}_{\alpha\alpha}$ and $\gamma^{IJ}_{\beta\beta}$ are naturally represented in the corresponding $\alpha$- and $\beta$-spin molecular-orbital spaces. Because the two orbital sets may differ in phase and spatial localization, these transition 1-RDMs cannot, in general, be combined directly when evaluating \cref{eq:soc_working_equation}. We therefore transform $\gamma^{IJ}_{\beta\beta}$ into the $\alpha$-spin molecular-orbital basis and use this common orbital representation for all quantities entering the SOC matrix elements.

Transition 1-RDMs between ADC states are required for all three ADC formulations considered in this work. For EE-ADC, construction of the SI Hamiltonian additionally requires the 1-RDM of the reference state and transition 1-RDMs connecting the reference state with the excited states. The latter are readily available from the EE-ADC calculation through the spectroscopic amplitudes $\mathbf{X}_\pm$ introduced in \cref{sec:theory:ADC}. 

The total spin quantum number $S$ of each ADC state is assigned from the expectation value $\langle \Psi_I \vert \hat{S}^{2} \vert \Psi_I \rangle$, evaluated as described in Refs.\@ \citenum{stahl:2022p044106}, \citenum{stahl:2024p204104}, and \citenum{stahl:2026p209901}. Specifically, for the $I$th ADC state $\ket{\Psi_I}$, we estimate
$S  =  \tfrac{1}{2}(-1 + \sqrt{1 + 4\langle \Psi_I \vert \hat{S}^{2} \vert \Psi_I \rangle})$.
For calculations involving open-shell reference wavefunctions, the resulting value may deviate from an exact spin quantum number because of spin contamination. We therefore assign $S$ by rounding the computed value to the nearest allowed half-integer or integer for states containing an odd or even number of electrons, respectively. The SI-ADC implementation issues a warning when the spin contamination exceeds a predefined threshold (0.01 by default), and states with substantial spin contamination can optionally be excluded from the state--interaction manifold.

The Wigner--Eckart construction further requires states with well-defined spin-projection quantum numbers $M$. 
While most ADC wavefunctions are eigenfunctions of the spin-projection operator $\hat{S}_z$ by construction, this property may be violated if two or more states are degenerate in energy. 
To ensure that the ADC wavefunctions are proper $\hat{S}_z$ eigenfunctions, we construct the matrix of spin projection operator in the basis of ADC states:
\begin{align}
	\label{eq:M_matrix}
	(S_z)_{IJ} \equiv \bra{\Psi_{I}}{\hat{S}_z}\ket{\Psi_{J}} = \frac{1}{2} \sum_{p} \left ( \gamma^{IJ}_{p\alpha p\alpha} - \gamma^{IJ}_{p\beta p\beta}\right )
\end{align}
Diagonalization of $\mathbf{S_z}$ yields a transformed set of wavefunctions $\ket{\tilde{\Psi}_I}$ with well-defined values of $M$. The corresponding eigenvectors are then used to transform the spin-resolved transition 1-RDMs into the basis of spin-projection eigenstates.

Finally, the states with the smallest positive common value of the spin-projection quantum number $\tilde{M}$ are selected to construct the Wigner 1-RDMs. These quantities are subsequently used in \cref{eq:soc_working_equation} to generate the SOC matrix elements for all required combinations of $S$, $S'$, $M$ and $M'$. The resulting state--interaction Hamiltonian is then diagonalized to obtain the final spin--orbit-coupled energies and states.

\subsection{Magnetic Properties: $g$-Tensor}
\label{sec:theory:magnetic}
Having established the SI framework within ADC for incorporating spin--orbit SOC effects, we now extend this formulation to the evaluation of the electronic $ g$-tensor, which characterizes the anisotropic interaction between the electron spin magnetic moment and an external magnetic field $\mathbf{B}$. This interaction is parametrized by the effective spin Hamiltonian,
\begin{align}
\hat{H}_{\mathrm{S}} = \mu_{B}\mathbf{B}^\mathrm{T} \mathbf{g}\mathbf{\tilde{S}},
\label{eq:SH}
\end{align}
where $\mu_{B}$ is the Bohr magneton\cite{kahn:1993p}, $\mathbf{B}$ denotes the external magnetic field, and $\mathbf{\tilde{S}}$ is the effective spin operator\cite{vancoillie:2007p1803,chibotaru:2012p064112} defined within the target-state manifold.

The electronic $g$-tensor is obtained by mapping the \emph{ab initio} Zeeman Hamiltonian onto the effective spin Hamiltonian. The \emph{ab initio} Zeeman Hamiltonian is given by
\begin{align}
	\label{eq:Hze}
	\hat{H}_{\mathrm{Ze}} = \mu_{B}\mathbf{B}^\mathrm{T} \bigl(\mathbf{L} + g_{e}\mathbf{S}\bigr) = \mu_{B}\mathbf{B}^\mathrm{T} \mathbf{J},
\end{align}
where $\mathbf{L} $ and $\mathbf{S} $ are the orbital and spin angular momentum operators, respectively, $\mathbf{J}=\mathbf{L}+g_{e}\mathbf{S}$ is the magnetic moment operator, and $g_e$ is the free-electron $g$-factor\cite{fan:2023p071801}.
By equating the matrix elements of the \emph{ab initio} Zeeman Hamiltonian and the effective spin Hamiltonian, the following working equations for the $g$-tensor are obtained\cite{chiang:2026p174117}:
\begin{align}
 \begin{aligned}
  & g_{kx} = \sqrt{\frac{2}{\tilde{S}}}\operatorname{Re}[ \mathbf{J}^{k}_{\tilde{S},\tilde{S}-1}] 
  \\ &
   g_{ky}  = -\sqrt{\frac{2}{\tilde{S}}}\operatorname{Im}[\mathbf{J}^{k}_{\tilde{S},\tilde{S}-1}  ] 
  \\ &
  { g_{kz} = \frac{1}{\tilde{S}} \mathbf{J}^{k}_{\tilde{S}\tilde{S}} }
 \end{aligned}  
\label{eq:g_working}
\end{align}
where $k\in\{x,y,z\}$
In this work, the magnetic moment operator $\mathbf{J}$ is evaluated within the SI-ADC framework, and its matrix elements are substituted into the working equations \eqref{eq:g_working} to obtain the electronic $g$-tensor. Note that the Zeeman Hamiltonian should, in principle, be transformed consistently with the decoupling transformation of the four-component Dirac equation to account for picture-change effects. However, previous studies have shown that neglecting these effects introduces only negligible errors in molecular $g$-tensor calculations.\cite{lan:2015p1750} Therefore, picture-change corrections to the Zeeman Hamiltonian are neglected in the present work.

\subsection{Limitations of the Single-Reference SI-ADC Approach}
\label{sec:theory:challenges}

Although state--interaction single-reference ADC (SI-ADC) provides an
efficient framework for incorporating spin--orbit coupling (SOC) and
electron correlation, its applicability is subject to three principal
limitations:

\begin{enumerate}
	
	\item \textit{Vanishing Clebsch--Gordan coefficients for certain
		spin projections.} The Wigner--Eckart reconstruction of SOC matrix elements in
	\cref{eq:soc_working_equation} relies on a nonzero Clebsch--Gordan
	(CG) coefficient for the chosen reference spin projection
	$\tilde{M}$. For $\tilde{M} = 0$, the coefficient
	$\langle S'0;1\,0|S0 \rangle$ vanishes when $S = S'$, preventing
	the corresponding Wigner 1-RDM from being extracted using
	\cref{eq:Wigner_rdm}. This limitation is particularly relevant
	for even-electron systems, where integer-spin states admit
	$M = 0$ components. When two or more states with the same
	nonzero spin are included in the SI manifold, using only their
	$M = 0$ components prevents the reconstruction of SOC matrix
	elements between these states, even though the corresponding
	couplings may be nonzero. Singlet states are excluded from this
	consideration, since two singlet states cannot couple through the
	spin--orbit Hamiltonian.
	
	This difficulty can be overcome by evaluating the Wigner 1-RDMs
	using states with a nonzero common spin projection
	$\tilde{M} > 0$, for which the relevant CG coefficient does
	not vanish. More generally, the reference spin projection can
	be selected separately for each pair of spin manifolds to
	ensure that the Wigner--Eckart reconstruction remains
	well-defined. In particular, this limitation does not arise
	for odd-electron states ($S = 1/2$, $S = 3/2$, $\ldots$), for which a nonzero spin
	projection (e.g., $\tilde{M} = 1/2$) is always available.
	
	\item \textit{Spin contamination in single-reference ADC states.}
	The Wigner--Eckart theorem assumes that the interacting states
	are eigenfunctions of both $\hat{S}^{2}$ and $\hat{S}_z$,
	with well-defined spin quantum numbers $S$ and $M$.
	However, as discussed in \cref{sec:theory:SIADC}, single-reference
	ADC calculations based on unrestricted
	Hartree--Fock (UHF) references may produce spin-contaminated
	states, particularly for open-shell systems.\cite{stahl:2022p044106,stahl:2024p204104}
	In such cases,
	the computed expectation value $\langle \hat{S}^{2} \rangle$
	deviates from the exact eigenvalue $S(S+1)$, indicating
	admixtures of states with different total spin.
	The resulting SOC matrix elements may therefore be affected
	by errors associated with the violation of spin symmetry.
	
	Several strategies can mitigate this limitation. Using a restricted
	open-shell Hartree--Fock (ROHF) reference can reduce spin
	contamination and improve the spin purity of the resulting
	states.\cite{stahl:2022p044106,stahl:2024p204104} 
	Alternatively, the spin expectation values can be
	monitored to identify and exclude strongly spin-contaminated
	states from the SI manifold (\cref{sec:theory:SIADC}). 
	For systems with substantial
	multireference character or persistent spin contamination,
	spin-adapted multireference ADC (MR-ADC) methods provide a
	more general route to constructing spin-pure correlated
	states suitable for the SI treatment of SOC.\cite{sokolov:2018p204113,chatterjee:2019p5908,chatterjee:2020p6343,mazin:2021p6152} 
	
	\item \textit{Incorrect description of orbitally degenerate
		reference states.} An additional limitation arises when 
	SI-ADC is applied to
	systems with orbitally degenerate reference states. 
	In the single-reference ADC($n$) methods,
	one component of the degenerate reference manifold is
	described by the $n$th-order M\o ller--Plesset perturbation
	theory (MP$n$), whereas the remaining components
	are obtained as excited states through the ADC
	eigenvalue problem. Because the reference and excited states
	are treated using different approximations, their energies
	do not preserve the degeneracy of the
	underlying spin-free Hamiltonian.
	This imbalance can introduce artificial splittings within
	the reference manifold, which may distort the subsequent
	state--interaction treatment of SOC, resulting in large errors
	in computed energy splittings and magnetic
	properties.
	
	A more appropriate treatment of such systems requires
	a framework that describes all components of the degenerate
	manifold on an equal footing. Multireference ADC, combined
	with a suitable multistate reference treatment, provides
	a route to addressing this limitation by treating the
	relevant electronic configurations within a common
	theoretical framework.\cite{sokolov:2018p204113,chatterjee:2019p5908,chatterjee:2020p6343,mazin:2021p6152} 
	In the present single-reference
	implementation, we avoid this difficulty by restricting
	SI-ADC calculations to systems with orbitally
	nondegenerate ground states.
\end{enumerate}

Despite these limitations, SI-ADC offers a versatile and
computationally efficient approach for describing SOC in
correlated electronic states. By combining spin-free ADC
energies and transition density matrices with a
state--interaction treatment of SOC, the framework enables
the calculation of spin--orbit-coupled electronic energies,
spectroscopic observables, and magnetic properties across
a broad range of molecular systems with orbitally nondegenerate 
ground states.
In the following, we demonstrate the accuracy and
capabilities of single-reference SI-ADC through systematic benchmarks
and a few representative applications.

\section{Computational Details}
\label{sec:computation}

All excitation energies and one-particle reduced density matrices
(1-RDMs) were computed using the single-reference ADC implementation
in the \textsc{PySCF} package.\cite{sun:2026p102502}
Unless otherwise specified, the ADC calculations employed unrestricted
Hartree--Fock (UHF) reference wavefunctions.
Scalar relativistic effects were included in all calculations
using the spin-free exact two-component one-electron (sf-X2C-1e)
Hamiltonian.\cite{liu:2009p031104,li:2012p154114}
Density fitting\cite{dunlap:2000p2113} was employed to reduce the
computational cost.
The subsequent state--interaction (SI) calculations, including
the transformation of electronic states into a basis with
well-defined spin-projection quantum numbers ($M$), evaluation
of spin--orbit coupling (SOC) matrix elements, and calculation
of electronic $g$-tensors, were performed using the
\textsc{Prism} code.\cite{MouraSokolov2025Prism} 
The required SOC integrals for the Breit--Pauli (BP) and
sf-X2C-1e+so-DKH1 (XDKH1) Hamiltonians were generated using
the \textsc{Socutils} package.\cite{Wang2022Socutils}

Throughout this work, we distinguish three approaches to incorporating
SOC within the ADC framework.
The state--interaction formulation investigated here is denoted
SI-ADC, with SI-BP-ADC and SI-XDKH1-ADC referring to calculations
employing the BP and XDKH1 spin--orbit Hamiltonians, respectively.
For comparison, we consider the perturbative SOC-ADC approach
developed in Ref.\@ \citenum{majumder:2025p2414}, in which
spin--orbit coupling is incorporated directly into the ADC effective Hamiltonian
and is treated consistently with dynamical correlation up to second order.
This approach is denoted as SOC-BP-ADC or SOC-XDKH1-ADC according to
the underlying spin--orbit Hamiltonian. 
The key distinction between these two approaches lies in how SOC is incorporated into the ADC framework. In SI-ADC, the spin-free ADC calculations are performed first, followed by the construction and diagonalization of an effective Hamiltonian that includes SOC within a selected state manifold. In contrast, SOC-ADC incorporates SOC directly into the ADC effective Hamiltonian, treating SOC and dynamical correlation within the same perturbative framework.
Finally, the variational treatment of SOC within ADC based on
the four-component (4c) Dirac--Coulomb Hamiltonian,
developed in Ref.\@ \citenum{chakraborty:2025p104106},
is referred to as 4c-ADC.

As discussed in \cref{sec:theory:SOC}, the SI-ADC treatment requires spin-resolved 1-RDMs to evaluate the matrix elements of the spin--orbit Hamiltonian.
For all methods except SI-EE-ADC(3), the 1-RDMs associated with ADC($n$) were evaluated consistently through $n$th order in perturbation theory.
For SI-ADC(2)-X, the 1-RDMs additionally included third-order contributions arising from the double-excitation sector of the ADC excited-state eigenvectors.
In the case of SI-EE-ADC(3), the 1-RDMs were evaluated at the SI-EE-ADC(2)-X level, consistent with our previous work.\cite{stahl:2024p204104}
The density matrix entering the construction of the SOMF Hamiltonian was obtained as an equal-weight average of the state-specific 1-RDMs over all states included in the SI calculation.

For the open-shell systems considered in the zero-field splitting
(ZFS) and $g$-tensor benchmarks
(\cref{sec:results_1,sec:results_2,sec:results_5,sec:results_6}),
the target $N$-electron states were accessed using different
ADC formulations.
Specifically, EA-ADC and IP-ADC calculations employed
closed-shell $(N-1)$- and $(N+1)$-electron reference
wavefunctions, respectively.
In contrast, EE-ADC calculations employed an $N$-electron
reference wavefunction, with the reference state
$\ket{\Psi^{N}_{0}}$ in \cref{eq:spectral_gf_EE}
explicitly included in the construction of the SOC
Hamiltonian matrix in \cref{eq:SI_SOC}.

In \cref{sec:results_1}, we assess the accuracy of SI-ADC
in predicting ZFS of $^2P$ and $^2\Pi$ states through
comparisons with the corresponding SOC-ADC results.
All calculations in this section employed the uncontracted
ANO-RCC-VTZP basis set.\cite{roos:2004p2851}
The comparison is extended to transition-metal atoms with
$^2D$ states in \cref{sec:results_2}, for which the
contracted X2C-TZVPall-2c basis set\cite{pollak:2017p3696}
was used.

To investigate the differences between SI-ADC and 4c-ADC,
we computed the zero-field splitting in ionized states of 
halogen oxide anions $\text{ClO}^-$, $\text{BrO}^-$,
and $\text{IO}^-$ using IP-ADC (\cref{sec:results_3}).
The Dyall.av3z basis set\cite{dyall:2002p335,dyall:2006p441}
was employed for the halogen atoms (Cl, Br, and I),
while the uncontracted aug-cc-pVTZ basis
set\cite{dunning:1989p1007,kendall:1992p6796}
was used for oxygen.

In \cref{sec:results_4}, we examine the performance of
SI-ADC for core-ionization energies, extending the assessment
of SOC effects beyond valence electronic states.
These calculations employed IP-ADC in conjunction with
the core--valence separation (CVS)
approximation\cite{cederbaum:1980p206,barth:1981p1038}.
The uncontracted ANO-RCC basis set was used to provide
a flexible description of core-hole relaxation.

The accuracy of SI-ADC for electronic $g$-tensors is
investigated in \cref{sec:results_5}.
Calculations for ZnH, CdH, and HgH used the ANO-RCC basis set.
In all other calculations, the def2-TZVP basis set was used.\cite{weigend:2005p3297}

Finally, \cref{sec:results_6} explores the applicability
of SI-ADC to heavy-element compounds, including the CeCl$_6^{3-}$ and UO$_2^{+}$ complexes
containing $f$-block elements.
The contracted ANO-RCC-VTZP basis set was used for these molecules.

Molecular geometries and the number of electronic states included
in each ADC calculation are provided in the Supporting Information.
Although the benchmark results discussed in \cref{sec:results} were computed with a fixed number of states,
we performed tests where we varied the size of state-interaction manifold. 
Tables S18 to S26 demonstrate that for most molecules, the SI-ADC results change very little once the first $\sim$10 excited states are included. The largest changes are observed for HgH, which shows significant differences in $g$-shift (within 10\%) for up to 20 states included in the calculation.

Table S17 reports the spin expectation values for the ground and excited states included in all SI-EE-ADC calculations. With the exception of the higher-lying states of Cu, Ag, and Au, which exhibit substantial spin contamination, the remaining states show only negligible deviations from the expected spin-pure values. In contrast, all SI-EA/IP-ADC calculations employed closed-shell reference states, and the resulting charged excited states were therefore free of spin contamination.

\section{Results and Discussion}
\label{sec:results}
\subsection{Zero-Field Splitting in Main-Group Atoms and Diatomic Molecules}	
\label{sec:results_1}

\begin{figure*}[t!]
	\centering
	\makebox[\textwidth][c]{%
		\includegraphics[width=1\textwidth]{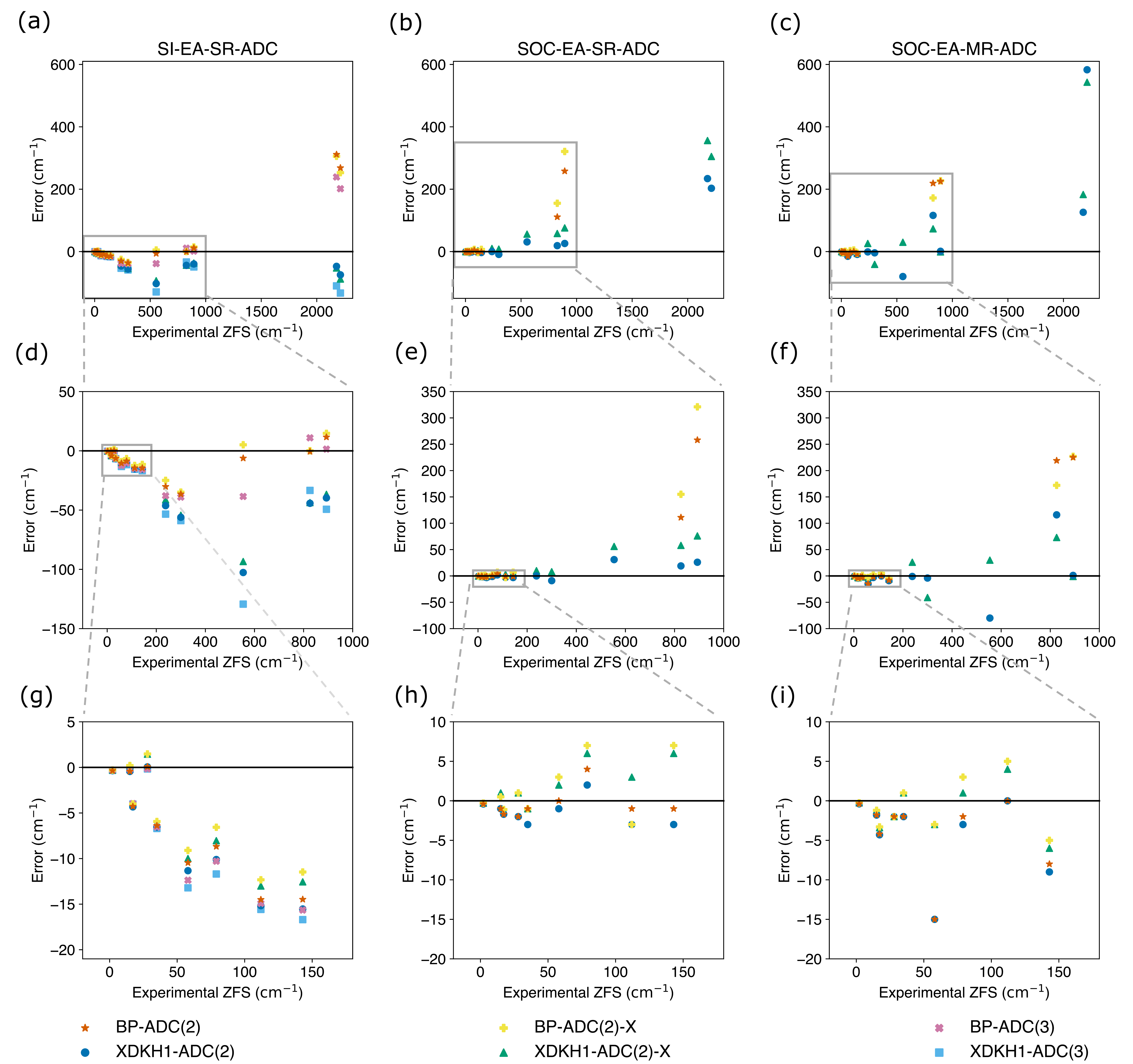}%
	}
	\caption{
		Errors in zero-field splittings (ZFS, \cm) computed using EA-ADC relative to experiment, plotted as a function of experimental ZFS values for atomic $^2P$ and molecular $^2\Pi$ states. The SOC-ADC results are taken from Ref.\@ \citenum{majumder:2025p2414}.
	}
	\label{fig:ea_p_zfs}
\end{figure*}

\begin{figure*}[t!]
	\centering
	\makebox[\textwidth][c]{%
		\includegraphics[width=1\textwidth]{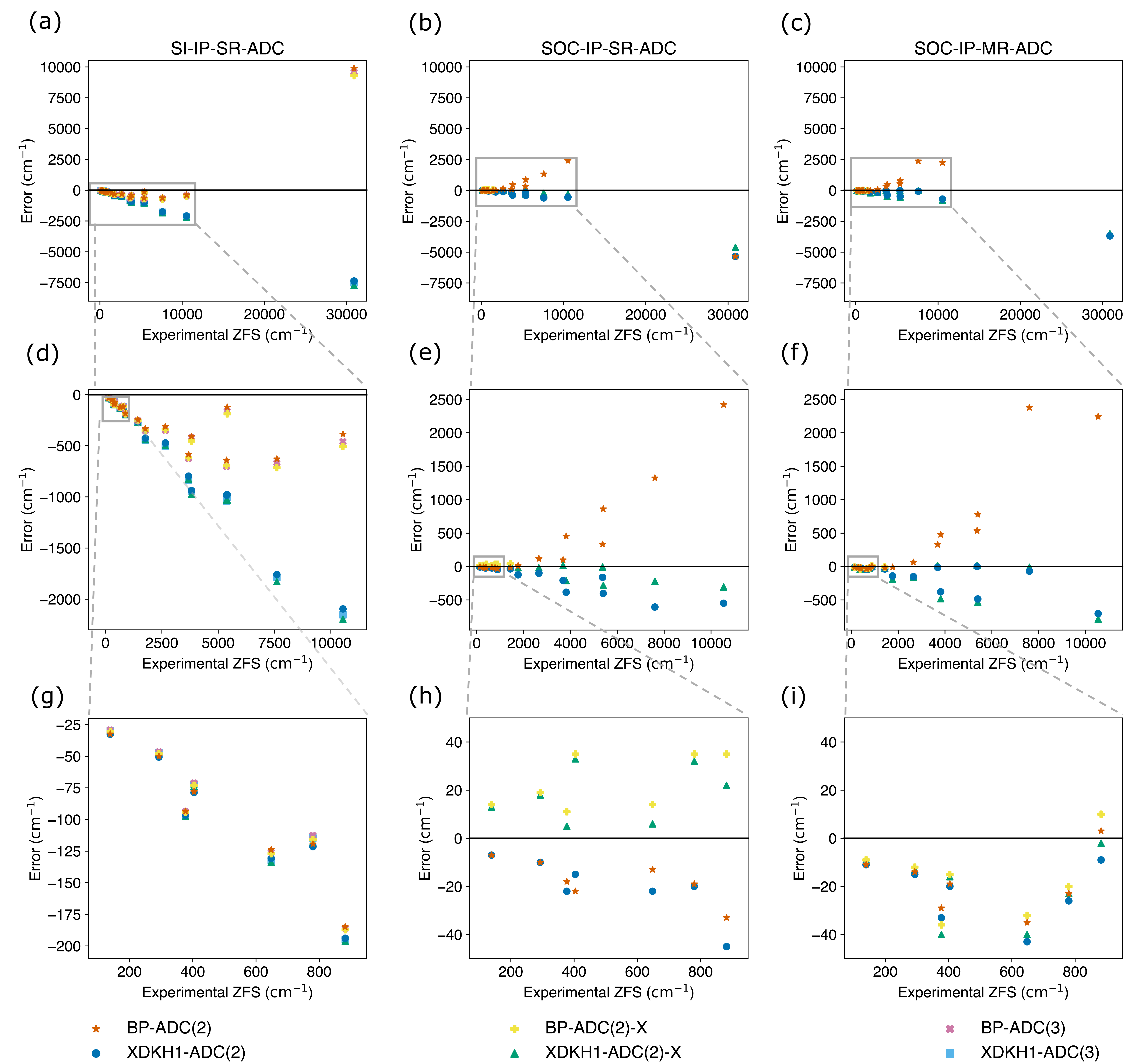}%
	}
	\caption{
		Errors in zero-field splittings (ZFS, \cm) computed using IP-ADC relative to experiment, plotted as a function of experimental ZFS values for atomic $^2P$ and molecular $^2\Pi$ states. The SOC-ADC results are taken from Ref.\@ \citenum{majumder:2025p2414}.
	}
	\label{fig:ip_p_zfs}
\end{figure*}

\begin{figure*}[t!]
	\centering
	\includegraphics[width=0.7\textwidth]{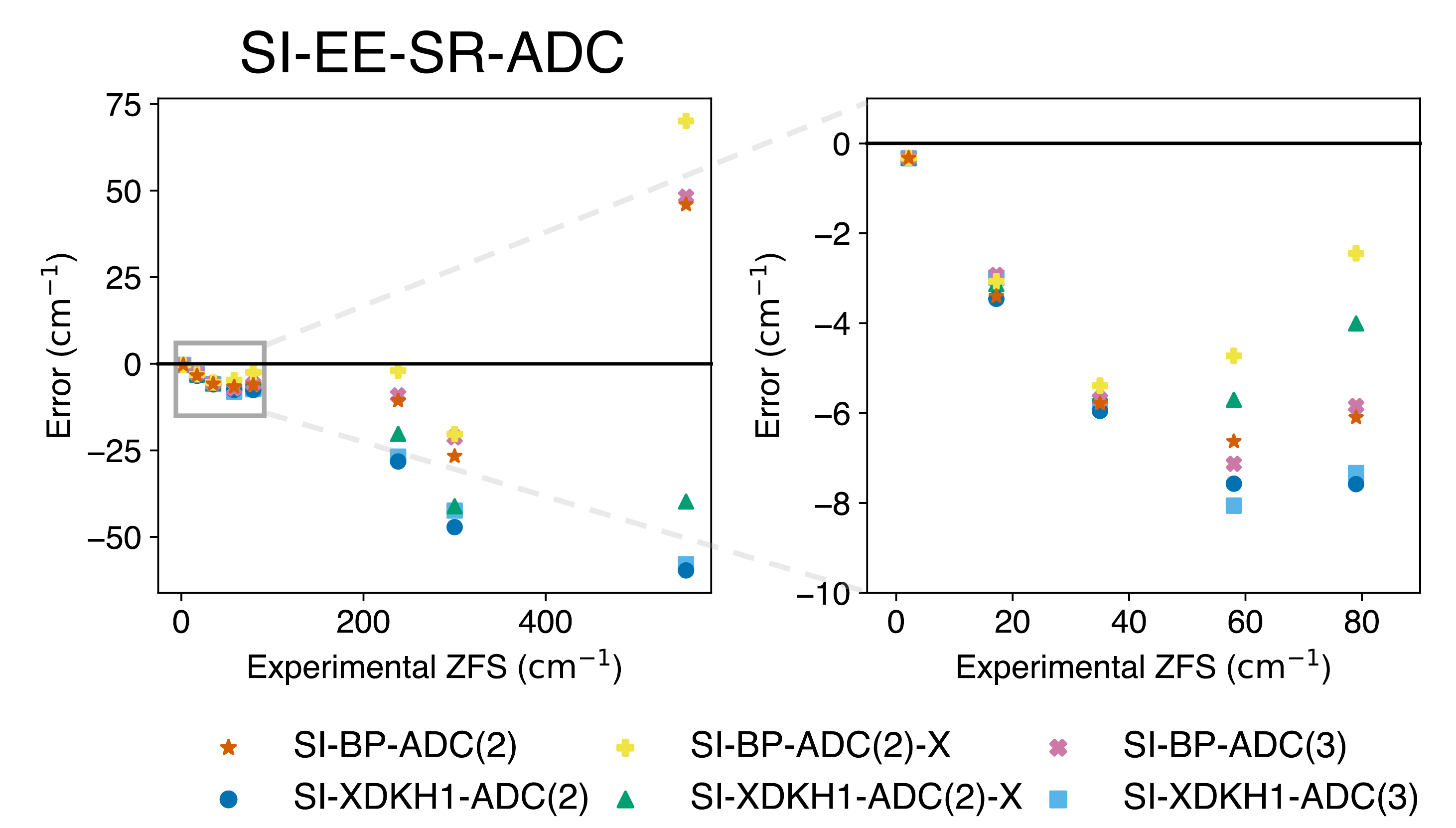}
	\caption{
		Errors in zero-field splittings (ZFS, \cm) computed using EE-ADC relative to experiment, plotted as a function of experimental ZFS values for atomic $^2P$ and molecular $^2\Pi$ states. 
	}
	\label{fig:ee_p_zfs}
\end{figure*}

To evaluate the accuracy of the single-reference state-interaction ADC
(SI-ADC) approach, we benchmark zero-field splittings (ZFS) for
a series of main-group atoms and diatomic molecules listed in the Supporting Information.
This benchmark set was previously investigated in
Ref.\@ \citenum{majumder:2025p2414} using the SOC-ADC approach,
which incorporates spin--orbit coupling (SOC) directly into the ADC effective Hamiltonian
and accounts for the interplay between SOC and dynamic electron
correlation through second order in perturbation theory.
In contrast, the SI-ADC approach employed here introduces SOC through
a state-interaction Hamiltonian constructed from spin-free ADC states,
with SOC matrix elements evaluated to first order.
By comparing SI-ADC with its single-reference and multireference
SOC-ADC counterparts (SOC-SR-ADC and SOC-MR-ADC, respectively),
we assess the importance of higher-order SOC contributions
and their interplay with electron correlation in determining ZFS.
For the EA- and IP-ADC calculations, the target $N$-electron
states are obtained from closed-shell singlet references with
$(N-1)$ and $(N+1)$ electrons, respectively,
whereas EE-ADC employs an open-shell $N$-electron reference.

\Cref{fig:ea_p_zfs,fig:ip_p_zfs,fig:ee_p_zfs} present the errors
in the computed ZFS values relative to experiment as a function
of experimental ZFS for the three SI-ADC formulations.
The corresponding numerical results for individual systems
are provided in Tables S1--S3.
Across the benchmark set, the Breit--Pauli (BP) and XDKH1
Hamiltonians generally yield similar results for systems
with small ZFS values, whereas their predictions diverge
as the magnitude of SOC increases.
In particular, the BP Hamiltonian tends to predict larger
ZFS values than XDKH1 for systems containing heavier elements.
This behavior reflects differences in the treatment of
relativistic effects by the two Hamiltonians.
The BP spin--orbit interaction contains an $r^{-3}$
dependence, where $r$ denotes the electron--nucleus distance,
which can lead to an overestimation of SOC for heavy elements
when the nonrelativistic form of the operator becomes
inadequate.
In contrast, the XDKH1 approach incorporates relativistic
effects through a unitary transformation of the Dirac
Hamiltonian, providing a more appropriate description of
the spin--orbit interaction in the vicinity of heavy
nuclei.\cite{dyall:2020p}
The resulting differences between BP and XDKH1 become
increasingly pronounced with increasing SOC strength,
although their relative accuracy also depends on the
underlying ADC formulation and the treatment of SOC (EA, IP, or EE).

\Cref{fig:ea_p_zfs} compares the performance of the
SI-EA-ADC and SOC-EA-ADC approaches.
For systems with experimental ZFS values in the
$500\text{--}1000~\mathrm{cm}^{-1}$ range,
SI-BP-EA-ADC generally provides better agreement with
experiment than SI-XDKH1-EA-ADC.
However, for systems with ZFS values exceeding
$2000~\mathrm{cm}^{-1}$, the BP Hamiltonian tends to
significantly overestimate the splitting, whereas
XDKH1 yields more accurate results.

Interestingly, the relative performance of the two
Hamiltonians differs between the SI-ADC and SOC-ADC
formulations.
Within SOC-EA-ADC, XDKH1 generally provides more accurate
ZFS values than BP as the strength of SOC increases.
Moreover, the direct incorporation of the BP spin--orbit
Hamiltonian into the ADC effective Hamiltonian can lead
to substantial errors and, in some cases, unphysical
results, including large negative excitation energies.\cite{majumder:2025p2414}
This behavior arises from the lack of a variational lower
bound in the resulting effective Hamiltonian, which can
lead to an overestimation of SOC effects.
In contrast, SI-BP-EA-ADC remains numerically well behaved
for the systems considered here and yields results
comparable to those obtained with SI-XDKH1-EA-ADC.
This unexpectedly favorable performance likely reflects
error cancellation, whereby the omission of higher-order
SOC contributions in the state-interaction treatment
partially compensates for the tendency of the BP Hamiltonian
to overestimate SOC effects.

Another notable feature of SI-EA-ADC is its weak
dependence on the ADC perturbation order.
The ADC(2), ADC(2)-X, and ADC(3) approximations
produce nearly identical ZFS values across the
benchmark set, indicating that higher-order corrections
to the spin-free electronic structure have a relatively
small effect on the predicted splittings for these
systems.
In terms of accuracy, SI-EA-ADC generally performs
less favorably than the corresponding SOC-SR-ADC and
SOC-MR-ADC approaches for systems with ZFS values
below $150~\mathrm{cm}^{-1}$.
However, for larger splittings, SI-BP-EA-ADC
achieves accuracy comparable to that of the
SOC-SR-ADC and SOC-MR-ADC approaches employing
the XDKH1 Hamiltonian.
These results demonstrate that the first-order
state-interaction treatment can provide an accurate
description of SOC-induced splittings over a broad
range of interaction strengths, despite neglecting
higher-order SOC contributions.

The SI-IP-ADC results in \cref{fig:ip_p_zfs}
exhibit several trends similar to those observed
for SI-EA-ADC.
In particular, the computed ZFS values show only
a weak dependence on the ADC perturbation order,
with ADC(2), ADC(2)-X, and ADC(3) producing
closely agreeing results. As the magnitude of ZFS increases, SI-BP-IP-ADC
generally provides better agreement with experiment
than SI-XDKH1-IP-ADC, with the notable exception
of the $\mathrm{Rn}^{+}$ cation.
This behavior contrasts with that of the corresponding
SOC-IP-ADC approach, for which the BP Hamiltonian
systematically overestimates the ZFS.
The improved performance of BP within SI-IP-ADC
is consistent with the favorable error cancellation
suggested by the EA-ADC results, although the magnitude
and effectiveness of this cancellation appear to depend
on the electronic structure of the system.

For systems with ZFS values below
$1000~\mathrm{cm}^{-1}$, SI-IP-ADC is generally
less accurate than the corresponding SOC-SR-ADC
and SOC-MR-ADC methods.
In contrast, for systems with ZFS values between
$1000$ and $11000~\mathrm{cm}^{-1}$,
SI-BP-IP-ADC achieves accuracy comparable
to that of the corresponding single-reference
and multireference SOC-IP-ADC approaches
employing the XDKH1 Hamiltonian.
Thus, as in the EA-ADC calculations, the
state-interaction treatment provides a competitive
description of relatively large SOC-induced
splittings without explicitly incorporating
higher-order SOC contributions into the
ADC effective Hamiltonian.

Finally, we investigate the performance of the
state-interaction approach within the EE-ADC
framework.
Unlike EA- and IP-ADC, EE-ADC employs an
$N$-electron reference wavefunction, and the
reference state $\ket{\Psi^{N}_{0}}$ must be
included in the state-interaction Hamiltonian
in \cref{eq:SI_SOC}.
As discussed in the limitations of the SI-ADC
approach (\cref{sec:theory:challenges}), single-reference EE-ADC does not
provide a balanced description of orbitally
degenerate ground-state manifolds.
Specifically, one component of the degenerate
manifold is described by the reference-state
perturbation theory, whereas the remaining
components are obtained as excited states
through the EE-ADC eigenvalue problem.
This imbalance can artificially lift the
ground-state degeneracy even in the absence
of SOC, compromising the subsequent evaluation
of SOC-induced splittings.
To avoid this difficulty, we restrict the
present SI-EE-ADC benchmark to ZFS
arising from excited states.

As shown in \cref{fig:ee_p_zfs},
the SI-EE-ADC results exhibit trends
similar to those observed for SI-IP-ADC.
The BP and XDKH1 Hamiltonians generally
provide comparable accuracy, although
BP yields more accurate ZFS values for
the 5th-period systems considered here.
Furthermore, the dependence of the computed
splittings on the ADC perturbation order
is relatively weak, with differences between
ADC(2), ADC(2)-X, and ADC(3) generally
smaller than those arising from the choice
of SOC Hamiltonian.

Overall, the EA-, IP-, and EE-ADC benchmarks
demonstrate that SI-ADC provides an accurate
description of SOC-induced splittings across
a broad range of systems and SOC strengths.
Its performance depends only weakly on the
ADC perturbation order but is more sensitive
to the choice of SOC Hamiltonian and the
underlying treatment of SOC.
Notably, the BP Hamiltonian often exhibits
unexpectedly favorable performance within
the state-interaction framework, suggesting
that error cancellation may compensate, in
part, for the limitations of the first-order
SOC treatment.
These findings establish SI-ADC as a
computationally efficient alternative to
approaches that incorporate SOC directly
into the ADC effective Hamiltonian.

\subsection{Zero-Field Splitting in Transition Metal Atoms}
\label{sec:results_2}

\begin{table*}[t!]
  \centering
\caption{Zero-field splitting parameters ($\mathrm{cm}^{-1}$) for the $^{2}D$ states of Sc, Y, and La atoms, calculated using single-reference SI-EA-ADC. Numbers in parentheses are the percentage errors relative to experimental results\cite{NIST_ASD,koseki:2019p2325,Sugar:1985}. For comparison, results from SOC-ADC\cite{majumder:2025p2414}, XDKH2-QDNEVPT2\cite{majumder:2024p4676},  X2C-MRCISD,\cite{hu:2020p2975} X2CSO-DSRG-MRPT2\cite{park:2025p12071},  SA-SO-DSRG-PT2\cite{wang:2026p8910}, X2C-DSRG-MRPT2,\cite{zhao:2026p7230} and DKH1-L-PDFT\cite{jangid:2026p318} are also included. All calculations employ the contracted X2C-TZVPall-2c basis set, except for DKH1-L-PDFT and X2C-SO-DSRG-MRPT2 calculations, which employ ANO-RCC-VTZP and X2C-TZVP basis sets, respectively.
}
    \begin{tabular}{lccc}
    \hline\hline
          & Sc    & Y     & La \\
          \hline
    \multicolumn{1}{l}{SI-BP-EA-ADC(2)} & 173 (2.7) & 587( 10.7) & 1410 (33.9) \\
	\multicolumn{1}{l}{SI-BP-EA-ADC(2)-X} & 178 (6.2) & 596 (12.5) & 1385 (31.6) \\
	\multicolumn{1}{l}{SI-BP-EA-ADC(3)} & 175 (4.3) & 580 (9.5) & 1297 (23.2) \\
	\multicolumn{1}{l}{SI-XDKH1-EA-ADC(2)} & 172 (2.5) & 577 (8.8) & 1344 (27.6) \\
	\multicolumn{1}{l}{SI-XDKH1-EA-ADC(2)-X} & 178( 6.0) & 586 (10.6) & 1322 (25.5) \\
	\multicolumn{1}{l}{SI-XDKH1-EA-ADC(3)} & 175 (4.1) & 571 (7.7) & 1243 (18.0) \\
	\hline
    \multicolumn{1}{l}{SOC-BP-EA-SR-ADC(2)} & 108 ($-$35.5) & 381 ($-$28.2) &  \\
    \multicolumn{1}{l}{SOC-BP-EA-SR-ADC(2)-X} & 131 ($-$22.0) & 450 ($-$15.1) &  \\
    \multicolumn{1}{l}{SOC-XDKH1-EA-SR-ADC(2)} & 110 ($-$34.7) & 392 ($-$26.1) & 987 ($-$6.2) \\
    \multicolumn{1}{l}{SOC-XDKH1-EA-SR-ADC(2)-X} & 132 ($-$21.3) & 457 ($-$13.7) & 1094 (3.9) \\
    \hline
    \multicolumn{1}{l}{SOC-BP-EA-MR-ADC(2)} & 109 ($-$35.3) & 385 ($-$27.4) & 973 ($-$7.6) \\
	\multicolumn{1}{l}{SOC-BP-EA-MR-ADC(2)-X} & 132 ($-$21.3) & 454 ($-$14.3) &  \\
	\multicolumn{1}{l}{SOC-XDKH1-EA-MR-ADC(2)} & 110 ($-$34.6) & 396 ($-$25.3) & 1002 ($-$4.9) \\
	\multicolumn{1}{l}{SOC-XDKH1-EA-MR-ADC(2)-X} & 133 ($-$20.7) & 461 ($-$12.9) & 1089 (3.4) \\
	\hline
    \multicolumn{1}{l}{XDKH2-QDNEVPT2} & 141 ($-$16.3) & 428 ($-$19.2) & 897 ($-$14.9) \\
	\multicolumn{1}{l}{X2C-MRCISD} & 186 (10.2) & 524 ($-$1.1) & 936 ($-$11.2) \\
	    X2CSO-DSRG-MRPT2 & 218 (29.6) & 540 (2.0) & 944 ($-$10.4) \\       
	     SA-SO(DKH1)-DSRG-PT2 & 143 ($-$15.2) & 452 ($-$14.8) & 949 ($-$9.9) \\    
	     SA-SO(X2C)-DSRG-PT2 & 153 ($-$9.2) & 472 ($-$11.0) & 964 ($-$8.4) \\
	     X2C-DSRG-MRPT2 & 153 ($-$9.1) & 504 ($-$4.9) & 981 ($-$6.8) \\ 
	    DKH1-L-PDFT & 167 ($-$0.6) & 492 ($-$7.2) & 838 ($-$20.4) \\
	\hline
    {Experiment} & {168} & {530} & {1053} \\
    \hline\hline
    \end{tabular}%
  \label{tab:ScYLa}%
\end{table*}%

\begin{table*}[t!]
  \centering
  \caption{Zero-field splitting parameters ($\mathrm{cm}^{-1}$) for the $^{2}D$ states of Cu, Ag, and Au atoms, calculated using single-reference SI-IP-ADC and SI-EE-ADC. Numbers in parentheses are the percentage errors relative to experimental results\cite{sansonetti:2005p1559,pickering:2001p181,sugar:1990p527}. For comparison, results from SOC-ADC are also included\cite{majumder:2025p2414}. All calculations employ the contracted X2C-TZVPall-2c basis set.
	}
    \begin{tabular}{lccc}
    \hline \hline
          & Cu    & Ag    & Au \\
       \hline
    \multicolumn{1}{l}{SI-BP-IP-ADC(2)} & 1827 ($-$10.6) & 4226 ($-$5.5) & 12927 (5.3) \\
	\multicolumn{1}{l}{SI-BP-IP-ADC(2)-X} & 1854 ($-$9.2) & 4212 ($-$5.8) & 12873 (4.9) \\
	\multicolumn{1}{l}{SI-BP-IP-ADC(3)} & 1808 ($-$11.5) & 4156 ($-$7.1) & 12683 (3.3) \\
	\multicolumn{1}{l}{SI-XDKH1-IP-ADC(2)} & 1820 ($-$10.9) & 4133 ($-$7.6) & 11826 ($-$3.6) \\
	\multicolumn{1}{l}{SI-XDKH1-IP-ADC(2)-X} & 1847 ($-$9.6) & 4120 ($-$7.9) & 11792 ($-$3.9) \\
	\multicolumn{1}{l}{SI-XDKH1-IP-ADC(3)} & 1801 ($-$11.8) & 4067 ($-$9.1) & 11645 ($-$5.1) \\
		\hline
	\multicolumn{1}{l}{SI-BP-EE-ADC(2)} & 1898 ($-$7.1) & 4336 ($-$3.0) & 13384 (9.0) \\
	\multicolumn{1}{l}{SI-BP-EE-ADC(2)-X} & 1902 ($-$6.9) & 4324 ($-$3.3) & 13349 (8.8) \\
	\multicolumn{1}{l}{SI-BP-EE-ADC(3)} & 1896 ($-$7.2) & 4274 ($-$4.4) & 13157 (7.2) \\
	\multicolumn{1}{l}{SI-XDKH1-EE-ADC(2)} & 1891 ($-$7.4) & 4242 ($-$5.1) & 12274 (0.0) \\
	\multicolumn{1}{l}{SI-XDKH1-EE-ADC(2)-X} & 1895 ($-$7.3) & 4229 ($-$5.4) & 12236 ($-$0.3) \\
	\multicolumn{1}{l}{SI-XDKH1-EE-ADC(3)} & 1889 ($-$7.5) & 4182 ($-$6.5) & 12085 ($-$1.5) \\
		\hline
    \multicolumn{1}{l}{SOC-BP-IP-SR-ADC(2)} & 1787 ($-$12.5) & 4071 ($-$9.0) & 11547 ($-$5.9) \\
	\multicolumn{1}{l}{SOC-BP-IP-SR-ADC(2)-X} & 2181 (6.8) & 4727 (5.7) &  \\
	\multicolumn{1}{l}{SOC-XDKH1-IP-SR-ADC(2)} & 1785 ($-$12.6) & 4027 ($-$9.9) & 11105 ($-$9.5) \\
	\multicolumn{1}{l}{SOC-XDKH1-IP-SR-ADC(2)-X} & 2177 (6.6) & 4659 (4.2) & 13030 (6.2) \\
	\hline
    \multicolumn{1}{l}{SOC-BP-IP-MR-ADC(2)} & 1927 ($-$5.7) & 4291 ($-$4.0) & 12109 ($-$1.3) \\
    \multicolumn{1}{l}{SOC-BP-IP-MR-ADC(2)-X} & 1984 ($-$2.9) & 4344 ($-$2.9) &  \\
    \multicolumn{1}{l}{SOC-XDKH1-IP-MR-ADC(2)} & 1925 ($-$5.8) & 4245 ($-$5.1) & 11602 ($-$5.5) \\
    \multicolumn{1}{l}{SOC-XDKH1-IP-MR-ADC(2)-X} & 2019 ($-$1.1) & 4292 ($-$4.0) & 11490 ($-$6.4) \\
    \hline
    {Experiment} & {2043} & {4472} & {12274} \\
    \hline \hline
    \end{tabular}%
  \label{tab:CuAgAu}%
\end{table*}%

Next, we extend our assessment of SI-ADC to the ZFS of 
transition-metal atoms with $\mathrm{d}^{1}$ (Sc, Y, and La)
and $\mathrm{d}^{9}$ (Cu, Ag, and Au) electronic configurations.
These systems provide an opportunity to examine the performance
of the state-interaction approach for open-shell transition-metal
states with different electronic configurations and increasing
SOC strengths.

\Cref{tab:ScYLa} presents the ZFS values computed using different
ADC approaches for the $\mathrm{d}^{1}$ atoms.
Previous studies reported relative errors exceeding 20\% for
the ZFS of Sc and Y obtained using SOC-EA-SR-ADC and
SOC-EA-MR-ADC.\cite{majumder:2025p2414}
In contrast, the SI-EA-ADC approach reduces these errors
to below 10\%, achieving accuracy comparable to that of
more computationally demanding methods, including
spin--orbit quasi-degenerate N-electron valence perturbation theory
(XDKH2-QDNEVPT2)\cite{majumder:2024p4676}  two-component
multireference configuration interaction
(X2C-MRCISD)\cite{hu:2020p2975}, {state-interaction and variational formulations of two-component second-order driven similarity renormalization group multireference perturbation theory (X2CSO-DSRG-MRPT2, SA-SO-DSRG-PT2 and X2C-DSRG-MRPT2),\cite{park:2025p12071,wang:2026p8910,zhao:2026p7230} and spin--orbit linearized pair-density functional theory (DKH1-L-PDFT).\cite{jangid:2026p318}}
The improved agreement with experiment suggests that the
first-order state-interaction treatment can provide a
more accurate description of the ZFS in Sc and Y than
the direct incorporation of SOC into the ADC effective
Hamiltonian.

This trend, however, does not extend uniformly across the
$\mathrm{d}^{1}$ series.
For La, the heaviest atom in this group, SI-EA-ADC exhibits
larger deviations from experiment than for Sc and Y, whereas
the SOC-ADC approach\cite{majumder:2025p2414} provides a
more accurate description.
This contrasting behavior suggests that the first-order
treatment of SOC within the state-interaction framework
may become less reliable for heavier elements, where
higher-order SOC contributions and their interplay with
electron correlation become increasingly important.

The SI-EA-ADC results also reveal systematic differences
associated with the choice of relativistic Hamiltonian
and the treatment of electron correlation.
For all three Group 3 atoms, the XDKH1 Hamiltonian
provides slightly more accurate ZFS values than BP.
Furthermore, SI-EA-ADC(3) consistently improves
agreement with experiment relative to the lower-order
ADC approximations.
Thus, unlike the relatively weak dependence on ADC
perturbation order observed for the main-group systems
in \cref{sec:results_1}, the higher-order treatment
of electron correlation produces a noticeable improvement
in the ZFS predictions for the $\mathrm{d}^{1}$ transition-metal atoms.

We next consider the $\mathrm{d}^{9}$ atoms Cu, Ag, and Au,
for which the computed ZFS values are summarized in
\cref{tab:CuAgAu}.
In contrast to the favorable performance observed for
the $\mathrm{d}^{1}$ systems, SI-IP-ADC and
SOC-IP-SR-ADC yield similar accuracy for the
$\mathrm{d}^{9}$ atoms, with both approaches generally
exhibiting larger deviations from experiment than
SOC-IP-MR-ADC.
These results indicate that the state-interaction
treatment does not systematically improve the ZFS
predictions over SOC-SR-ADC for this electronic
configuration.
Instead, the improved performance of SOC-IP-MR-ADC
highlights the importance of high-order
electron correlation effects for these systems.

The relative accuracy of the BP and XDKH1 Hamiltonians
also varies across the $\mathrm{d}^{9}$ series.
For Cu and Ag, BP yields slightly more accurate ZFS
values than XDKH1 when compared at the same ADC
perturbation order and within the same excitation
framework (EE or IP).
This behavior is consistent with the trends observed
for several main-group systems in \cref{sec:results_1},
where the BP Hamiltonian also exhibited unexpectedly
favorable performance.
Such agreement may arise from fortuitous error
cancellation between the approximations in the BP
Hamiltonian and the underlying electronic structure
treatment.

For Au, however, the trend is reversed: XDKH1
provides more accurate ZFS values than BP at the
same ADC perturbation order and within the same
excitation framework.
This result is consistent with the increasing
importance of higher-order relativistic effects
for heavier elements, for which the BP Hamiltonian
becomes less reliable.

Overall, the transition-metal benchmarks reveal
a pronounced dependence of SI-ADC accuracy on
the electronic configuration and the level of approximation 
due to the interplay of high-order electron correlation 
and relativistic effects in these systems.
For the $\mathrm{d}^{1}$ atoms, SI-EA-ADC
provides an accurate description of ZFS,
achieving agreement with experiment comparable
to that of higher-level approaches such as
SOC-MR-ADC, QDNEVPT2, and X2C-MRCISD.
In contrast, for the $\mathrm{d}^{9}$ atoms,
the SI-IP-ADC and SI-EE-ADC formulations
show no systematic improvement over their
single-reference SOC-ADC counterparts, while
SOC-IP-MR-ADC generally provides more accurate
results.
The choice of relativistic Hamiltonian introduces
an additional system-dependent effect, with
BP yielding slightly better agreement for Cu
and Ag and XDKH1 becoming more accurate for Au.

\subsection{Comparison of State--Interaction and Four-Component ADC Approaches}
\label{sec:results_3}
\begin{table*}[t!]
  \centering
\caption{
	Zero-field splitting energies ($\mathrm{cm}^{-1}$) for the energy levels of $\mathrm{ClO}$, $\mathrm{BrO}$, and $\mathrm{IO}$, calculated using single-reference SI-IP-ADC and SI-EE-ADC.
	Numbers in parentheses are the percentage errors relative to experimental results.\cite{gilles:1992p8012} 
	For comparison, results from 4c-IP-ADC and 4c-IP-EOM-CCSD in Ref.\citenum{chakraborty:2025p104106} are also included. 
	The contracted Dyall.av3z basis set is adopted for the halogen atoms (Cl, Br, and I), whereas the uncontracted aug-cc-pVTZ basis set is employed for the oxygen atom.
	}
    \begin{tabular}{lccc}
    \hline \hline
          & ClO & BrO & IO \\
    \hline
    SI-BP-IP-ADC(2) & 199.4 ($-$37.7) & 489.8 ($-$52.2) & 964.6 ($-$53.9) \\
    SI-BP-IP-ADC(2)-X & 212.8 ($-$33.5) & 554.0 ($-$45.9) & 1241.7 ($-$40.6) \\
    SI-BP-IP-ADC(3) & 264.9 ($-$17.3) & 927.2 ($-$9.5) & 2703.2 (29.3) \\
    SI-XDKH1-IP-ADC(2) & 198.0 ($-$38.2) & 463.5 ($-$54.7) & 825.1($-$60.5) \\
    SI-XDKH1-IP-ADC(2)-X & 211.2 ($-$34.0) & 523.2 ($-$48.9) & 1056.1 ($-$49.5) \\
    SI-XDKH1-IP-ADC(3) & 262.4 ($-$18.0) & 869.5 ($-$15.1) & 2272.9 (8.7) \\
    \hline
    4c-IP-ADC(2) & 251.6 ($-$21.4) & 613.8 ($-$40.1) & 1048.5 ($-$49.9) \\
	4c-IP-ADC(3) & 405.7 (26.7) & 1331.6 (30.0) & 3124.6 (49.4) \\
    4c-IP-EOM-CCSD & 340.4 (6.3) & 1043.7 (1.9) & 2238.9 (7.1) \\
	\hline
    {Experiment} & {320.2} & {1024.3} & {2091.4} \\
       \hline \hline
    \end{tabular}%
  \label{tab:4c-ADC}%
\end{table*}%

Having compared the state-interaction and effective-Hamiltonian
treatments of spin--orbit coupling (SOC) within the ADC
framework, we next benchmark SI-ADC against a fully relativistic
formulation.
Specifically, we consider the variational four-component ADC
(4c-ADC) method developed by Chakraborty and
co-workers,\cite{chakraborty:2025p104106} which employs a
Dirac--Hartree--Fock reference based on the Dirac--Coulomb
Hamiltonian and incorporates relativistic effects directly
throughout the correlated electronic structure calculation.
This rigorous treatment comes at a substantially increased
computational cost: 4c-ADC(3) calculations have been reported
to be up to 32 times more expensive than their nonrelativistic
counterparts.\cite{chakraborty:2026p3971}
In contrast, SI-ADC introduces SOC only after the spin-free
ADC calculation through state interaction, such that the
overall cost of SI-ADC(3) remains comparable to that of
spin-free ADC(3).
The comparison therefore provides a stringent assessment of
the accuracy attainable with the computationally inexpensive
SI treatment relative to a variational four-component
relativistic approach at different ADC perturbation orders.

\Cref{tab:4c-ADC} compares the ZFS values of the
neutral halogen monoxide radicals ClO, BrO, and IO,
obtained from IP-ADC calculations employing their
corresponding closed-shell anions as reference states.
At second order, 4c-ADC(2) provides more accurate
ZFS values than either SI-BP-IP-ADC(2) or
SI-XDKH1-IP-ADC(2), with relative errors
4 to 16 \% lower than those of the
corresponding SI-ADC approaches.
This result indicates that the fully relativistic
treatment offers an advantage over the first-order
state-interaction approximation at the ADC(2) level
for the systems considered here.

Interestingly, the relative performance of the
two approaches is reversed at third order.
While 4c-ADC(3) exhibits a substantial deterioration
in accuracy relative to 4c-ADC(2), SI-ADC(3)
provides more accurate ZFS values than its
four-component counterpart.
The improvement is particularly pronounced for IO,
where SI-XDKH1-IP-ADC(3) yields a relative error
of only 8.7\%, approaching the accuracy of the
considerably more computationally demanding
4c-IP-EOM-CCSD method, which exhibits an error
of 7.1\%.

The contrasting performance of the second- and
third-order methods can be understood by examining
the dependence of the computed ZFS values on the
ADC perturbation order.
In both the SI-ADC and 4c-ADC frameworks,
the predicted splittings increase upon going
from ADC(2) to ADC(3).
However, the magnitude of this increase differs
substantially between the two approaches.
For 4c-ADC, the third-order correction produces
a pronounced increase in the ZFS, leading to
a substantial overestimation of the experimental
values.
In contrast, the corresponding increase in
SI-ADC is considerably smaller, resulting in
more accurate predictions at third order.

Finally, we examine the influence of the
SOC Hamiltonian within the SI-ADC framework.
At a given ADC perturbation order, the BP
Hamiltonian generally provides more accurate
ZFS values than XDKH1 for the systems considered
here, with the notable exception of IO at
the ADC(3) level.
This behavior is consistent with the trends
observed in the preceding sections, where
the BP Hamiltonian frequently exhibited
unexpectedly favorable agreement with experiment
despite its known limitations for heavier
elements.
The results further suggest that favorable
error cancellation can compensate for
deficiencies in the BP approximation,
although this effect is neither systematic
nor guaranteed to persist with increasing
SOC strength.

Overall, these benchmarks demonstrate that
SI-ADC can provide ZFS predictions comparable
to those obtained using fully relativistic
four-component methods, despite employing
a substantially simpler treatment of SOC.
The particularly favorable performance of
SI-ADC(3) illustrates the potential of the
state-interaction framework for accurately
describing SOC-induced energy splittings
when combined with an appropriate treatment
of electron correlation.

\subsection{L-edge Core-Ionization Energies}
\label{sec:results_4}

\begin{figure*}[t!]
	\centering
	\includegraphics[width=\textwidth]{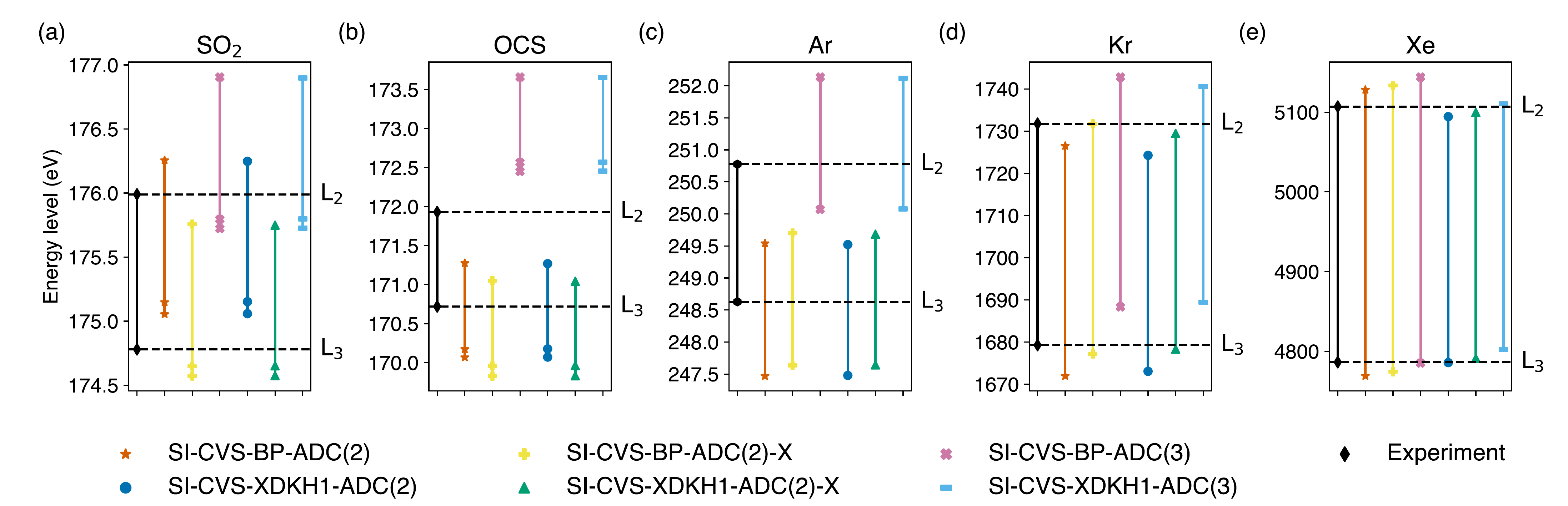}
	\caption{
		L$_{2,3}$-edge core ionization energies (eV) computed using SI-ADC methods. 
		The dashed horizontal line represents the experimental results.\cite{king:1977p2479,dragoun:2004p833,breinig:1980p520,coville:1995p21}
	}
	\label{fig:cvs_L23}
\end{figure*}

We next assess the performance of SI-ADC for calculating core-ionization
energies, which can be directly compared with X-ray photoelectron
spectroscopy (XPS) measurements.\cite{greczynski:2020p100591}
To efficiently access core-ionized states, we employ the core--valence
separation (CVS) approximation,\cite{cederbaum:1980p206,barth:1981p1038}
which neglects their coupling to the valence-ionized manifold.
\Cref{fig:cvs_L23} presents the L$_{2,3}$ core-ionization energies
of the noble-gas atoms Ar, Kr, and Xe and the molecules SO$_2$
and OCS computed using SI-CVS-IP-ADC.
The corresponding numerical results are reported in Table S4.

For the noble-gas atoms, the absolute errors in the calculated
L$_2$ and L$_3$ ionization energies increase substantially
from Ar to Xe.
Despite this deterioration in the absolute energies, the
relative errors in the spin--orbit splittings remain below
15\% for all three atoms.
This favorable performance likely reflects the large SOC
experienced by core electrons, which reduces the relative
sensitivity of the L$_{2,3}$ splitting to errors in the
treatment of electron correlation.
The choice of SOC Hamiltonian also affects the accuracy
of the predicted splittings.
While BP provides better agreement with experiment for Ar,
XDKH1 yields more accurate results for Kr and Xe,
consistent with the increasing importance of relativistic
effects for heavier elements.
In contrast, the computed splittings exhibit only a weak
dependence on the ADC perturbation order.

The molecular benchmarks for SO$_2$ and OCS further
demonstrate the applicability of SI-CVS-IP-ADC to core-level
spectroscopy.
Unlike the atomic systems, the nonspherical molecular
environment lifts the degeneracy of the L$_3$ manifold,
giving rise to additional, relatively small energy splittings.
The calculated L$_2$ and L$_3$ core-ionization energies
deviate from experiment by less than 2 eV, while the
relative errors in the corresponding spin--orbit splittings
remain below 5\%.
For SO$_2$, BP provides better agreement with experiment
than XDKH1, suggesting that favorable error cancellation
may contribute to its performance for this system.

Although the predicted spin--orbit splittings are relatively
insensitive to the ADC perturbation order, the absolute
core-ionization energies exhibit a more pronounced dependence.
In particular, SI-CVS-IP-ADC(3) generally produces larger
errors than SI-CVS-IP-ADC(2) and SI-CVS-IP-ADC(2)-X.
Similar behavior has been reported for CVS-EE-ADC and
CVS-IP-MR-ADC in Refs.\@ \citenum{wenzel:2015p214104}
and \citenum{demoura:2022p4769}, respectively.
As discussed in Ref.\@ \citenum{wenzel:2015p214104},
this trend can be attributed to the delicate balance
between orbital relaxation and polarization effects
in core-ionized states.
At second order, favorable error cancellation between
these contributions can yield accurate core-ionization
energies, whereas the inclusion of third-order corrections
may disrupt this balance and increase the overall error.
The contrasting sensitivity of the absolute ionization
energies and spin--orbit splittings to the ADC perturbation
order highlights the ability of SI-CVS-IP-ADC to capture
core-level SOC effects even when systematic errors remain
in the underlying core-ionization energies.

\subsection{Magnetic $g$-tensors}
\label{sec:results_5}

\begin{figure*}[t!]
	\centering
	\includegraphics[width=\textwidth]{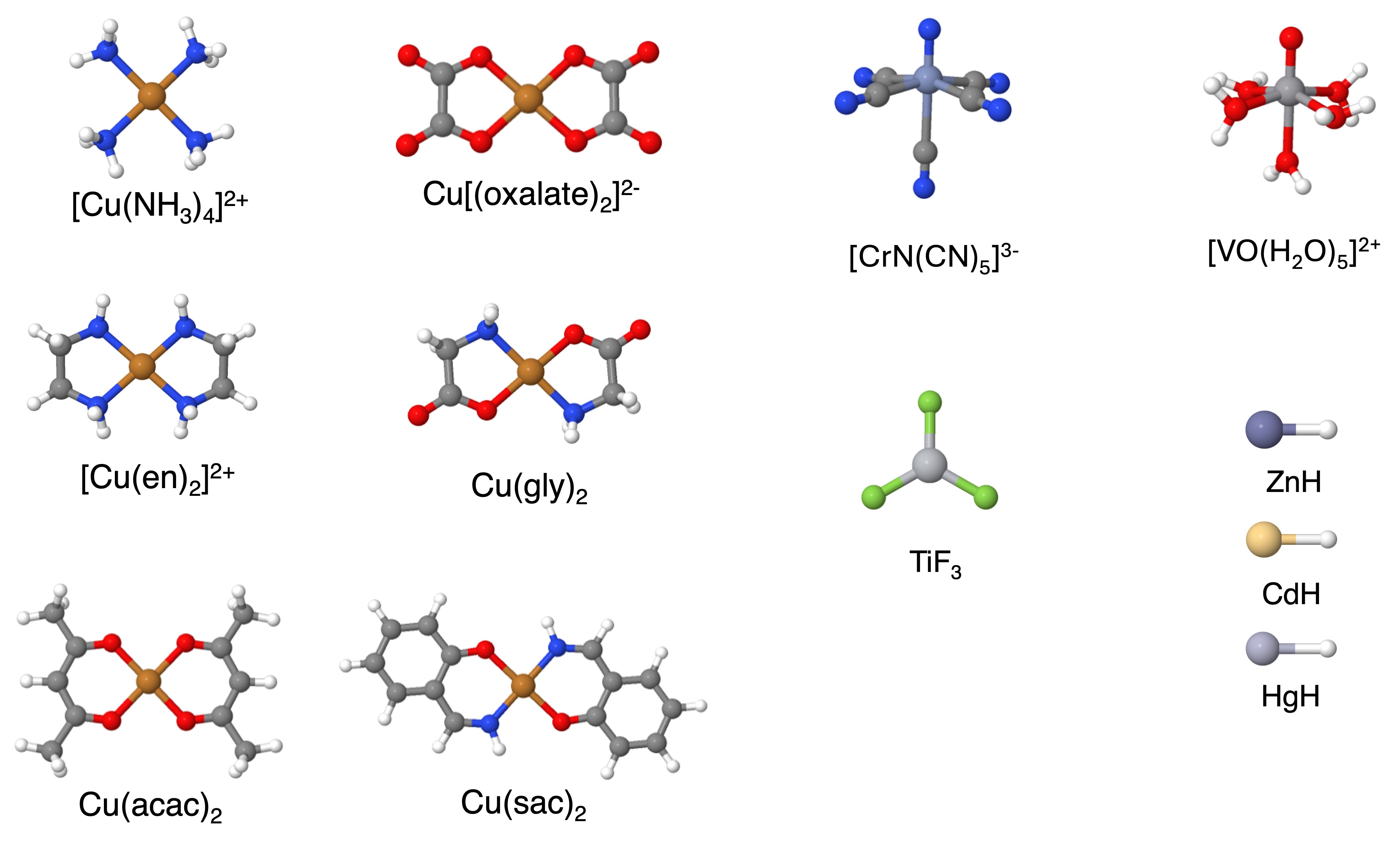}
	\caption{Molecular benchmark set for the $g$-tensor calculations (\cref{sec:results_5}). 
	}
	\label{fig:gtensor_molecules}
\end{figure*}

\begin{figure*}[t!]
	\centering
	\includegraphics[width=\textwidth]{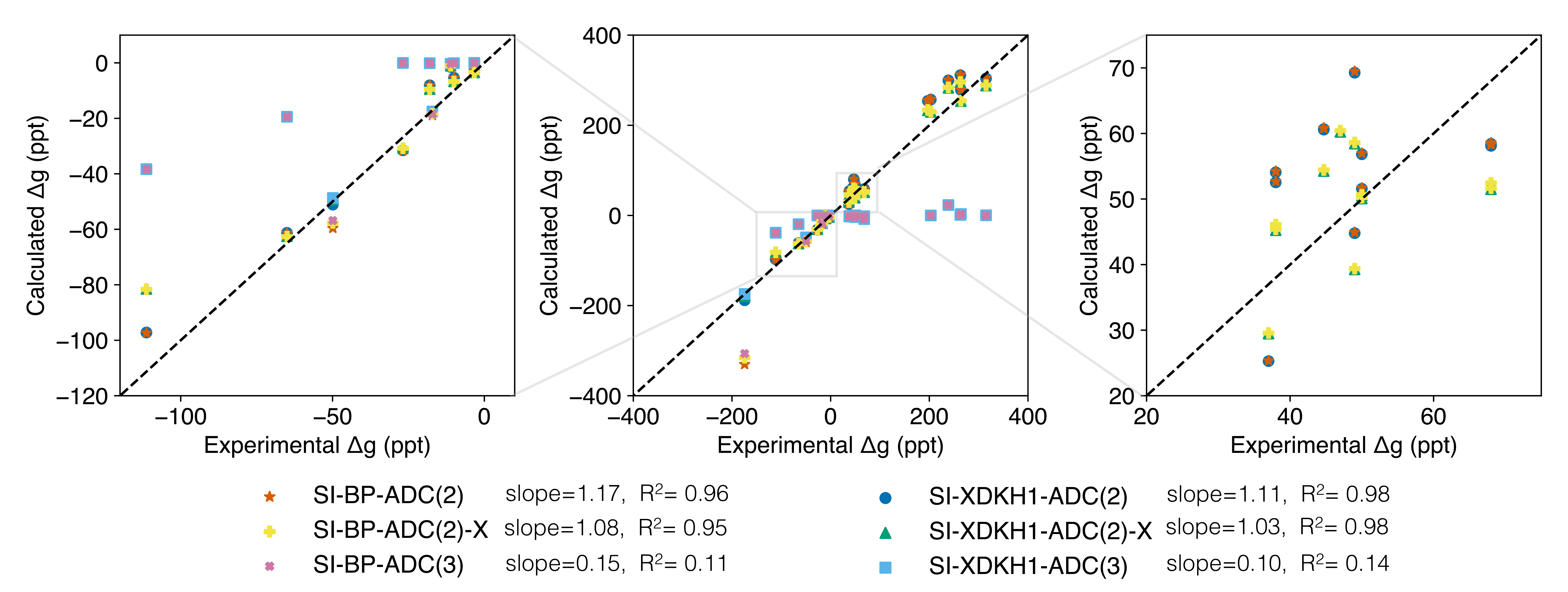}
	\caption{Correlation between experimental and SI-EA/IP-ADC-computed $g$-shifts (in parts per thousand, ppt) for molecules in \cref{fig:gtensor_molecules}.
	The results for molecules with negative experimental $g$-shifts were computed using SI-EA-ADC, while for molecules with positive experimental $g$-shifts the calculations were performed using SI-IP-ADC. 
	See Tables S5 to S9 for data on individual molecules.
	}
	\label{fig:large_ip_gtensor}
\end{figure*}

We next assess the accuracy of SI-ADC for molecular magnetic
properties, focusing on electronic $g$-tensors.
Our benchmark set spans systems ranging from diatomic molecules
to transition-metal complexes, which structures are shown in
\cref{fig:gtensor_molecules}.
The $g$-tensors of the Cu complexes were computed using SI-IP-ADC,
whereas the remaining systems were investigated using SI-EA-ADC.
For selected molecules, additional calculations were performed
using SI-EE-ADC.
All numerical results are provided in Tables S5--S9 of the
Supporting Information.

\Cref{fig:large_ip_gtensor} compares the calculated and experimental
$g$-shifts, defined relative to the free-electron value ($g_e$)
and expressed in parts per thousand
($\Delta g_i = (g_i-g_e)\times 10^3$, ppt).
The second-order SI-ADC methods exhibit a good agreement
with experiment across the benchmark set.
For SI-BP-ADC(2) and SI-BP-ADC(2)-X, linear regression yields
slopes of 1.17 and 1.08, respectively, with corresponding
$R^2$ values of 0.96 and 0.95.
The XDKH1-based methods provide slightly closer agreement
with experiment, with slopes of 1.11 and 1.03
and $R^2$ values of 0.98 for SI-XDKH1-ADC(2) and
SI-XDKH1-ADC(2)-X, respectively.
These results demonstrate that both second-order approximations
reproduce the experimental $g$-shifts over a broad range
of magnitudes and signs, with ADC(2)-X generally reducing
the systematic deviations observed at the ADC(2) level.

In contrast, the SI-ADC(3) results exhibit
substantially larger deviations from experiment.
The corresponding regression slopes decrease to 0.15
and 0.10 for BP and XDKH1, respectively, with $R^2$
values of only 0.11 and 0.13.
As illustrated in \cref{fig:large_ip_gtensor}, this
deterioration is particularly pronounced for systems
with large experimental $g$-shifts, for which ADC(3)
often predicts values much closer to the free-electron
limit.
A closer examination of the individual results in
Tables S5--S9 reveals that SI-ADC(3) yields severely underestimated
or unphysical $g$-shifts for all transition-metal
compounds shown in \cref{fig:gtensor_molecules} except ZnH, CdH, and HgH.

To understand the origin of SI-ADC(3) errors, we
analyzed the underlying spin-free (non-SI) EA- and IP-ADC(3)
results for the transition-metal complexes, which exhibit 
large discrepancies in $g$-shifts relative to experiment.
As shown in Tables S13 and S14 of the Supporting Information,
ADC(3) substantially overestimates the energies of
ionized and electron-attached states for all of these molecules, whereas the
corresponding second-order approximations provide
more accurate results.
This behavior is consistent with previous studies
demonstrating that single-reference ADC(3) can perform
considerably worse than ADC(2) for moderately and
strongly correlated systems, including transition-metal
complexes.\cite{chatterjee:2019p5908,chatterjee:2020p6343,mazin:2021p6152}
Because the electronic $g$-tensor depends on
SOC-induced mixing between electronic states, errors
in their relative energies can substantially affect
the calculated magnetic properties.
In particular, an overestimation of the energy
separations between interacting states suppresses
their SOC-induced mixing, potentially leading to
the severely underestimated $g$-shifts observed
in \cref{fig:large_ip_gtensor}.
The poor performance of SI-ADC(3) for these complexes
therefore originates primarily from deficiencies
in the underlying spin-free electronic structure
description, rather than from the state-interaction
treatment of SOC itself.

Overall, our benchmark demonstrates that second-order
SI-ADC provides an efficient and broadly applicable
framework for predicting molecular $g$-tensors.
XDKH1 generally improves accuracy for systems
containing heavy $p$-shell elements, whereas BP and
XDKH1 exhibit comparable performance for the
transition-metal complexes considered here.
The pronounced deterioration of SI-ADC(3) highlights
the importance of an accurate description of the
underlying spin-free electronic states for reliable
predictions of magnetic properties.
For more challenging open-shell systems, multireference
ADC may provide a promising route to improving the
description of electron correlation and extending
the applicability of the state-interaction framework.\cite{sokolov:2018p204113,chatterjee:2019p5908,chatterjee:2020p6343,mazin:2021p6152}

\subsection{Applications to Heavy-Element Compounds}
\label{sec:results_6}

\begin{table*}[t!]
	\centering
	\caption{Energy levels (cm$^{-1}$) of UO$_{2}^{+}$ computed using single-reference SI-EA-ADC and  SI-EE-ADC with the ANO-RCC-VTZP basis set. 
	The experimental results are from Ref. \citenum{merritt:2008p084304}.
	}
	 \resizebox{\textwidth}{!}{
     \begin{tabular}{cccccccc}   
     	\hline \hline 
     	Electronic state & SI-BP- & SI-BP- & SI-BP- & SI-XDKH1- & SI-XDKH1- & SI-XDKH1- & Experiment \\
     	(EA/EE)- & ADC(2) & ADC(2)-X & ADC(3) & ADC(2) & ADC(2)-X & ADC(3) &  \\
     	\hline    
     	$^{2}\Phi_{5/2u}$ (EA) & 0     & 0     & 0     & 0     & 0     & 0     &  \\
        $^{2}\Delta_{3/2u}$ (EA)  & 2745  & 3283  & 5896  & 2658  & 3179  & 6146  & 2658 \\
        $^{2}\Phi_{7/2u}$ (EA) & 7290  & 6942  & 9556  & 7049  & 6668  & 9306  &  \\
        $^{2}\Delta_{5/2u}$ (EA) & 7885  & 8105  & 10417 & 7658  & 7852  & 10910 &  \\   
        \hline  $^{2}\Phi_{5/2u}$ (EE)& 0     & 0     & 0     & 0     & 0     & 0     &  \\
        $^{2}\Delta_{3/2u}$ (EE)& 3117  & 2878  & 2290  & 3060  & 2805  & 2235  & 2658 \\
        $^{2}\Phi_{7/2u}$ (EE)& 6565  & 6567  & 6361  & 6344  & 6310  & 6136  &  \\
        $^{2}\Delta_{5/2u}$ (EE) & 7921  & 7746  & 7201  & 7742  & 7530  & 7067  &  \\    
     	 \hline \hline
     	\end{tabular}%
}
	\label{tab:UO2}%
\end{table*}%

\begin{figure*}[t!]
	\centering
	\includegraphics[width=\textwidth]{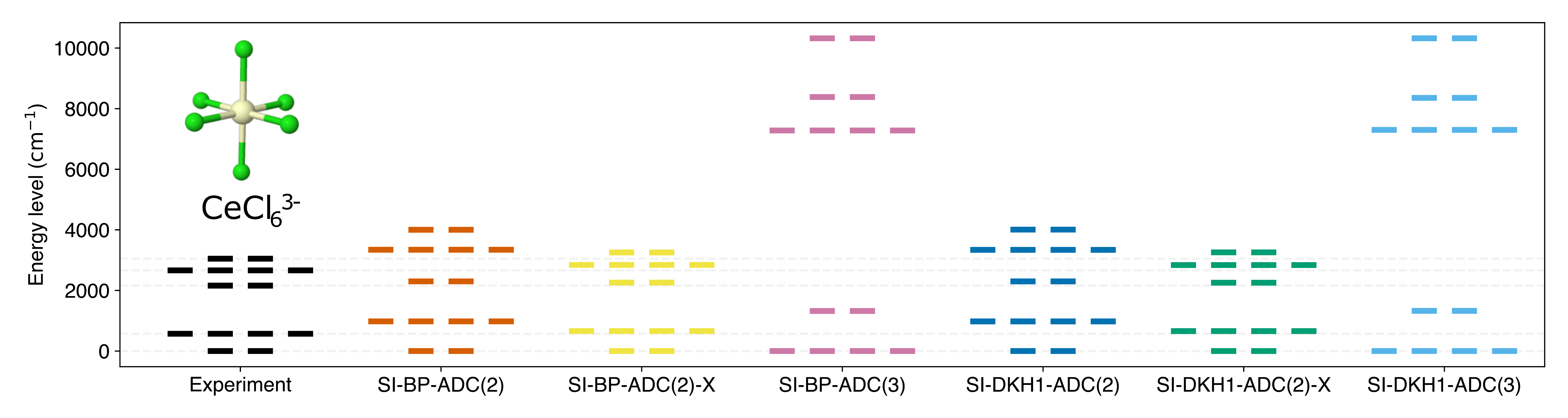}
	\caption{Energy levels (cm$^{-1}$) of CeCl$_{6}^{3-}$ computed using single-reference SI-EA-ADC with the ANO-RCC-VTZP basis set.
		The experimental results are from Ref. \citenum{jangid:2026p318}. }
	\label{fig:CeCl6}%
\end{figure*}

Finally, we benchmark the SI-ADC methods for $f$-block elements.
Extending SI-ADC to heavy-element compounds provides a stringent
test of the state-interaction framework, as the strong SOC and
complex electronic structure of $f$-block systems challenge
both the perturbative treatment of relativistic effects and
the underlying single-reference description of electron correlation.

We first consider UO$_2^+$, for which the energy levels computed
using SI-EA-ADC and SI-EE-ADC are presented in \cref{tab:UO2}.
For both formulations, the calculated energy levels are
considerably more sensitive to the ADC perturbation order
than to the choice of SOC Hamiltonian.
Among the SI-EA-ADC methods, ADC(2) provides the closest
agreement with the experimental $^2\Delta_{3/2u}$ excitation
energy, with SI-XDKH1-EA-ADC(2) reproducing the experimental
value of $2658~\mathrm{cm}^{-1}$.
In contrast, ADC(2)-X and ADC(3) progressively overestimate
this splitting, with ADC(3) also predicting substantially
higher energies for the remaining excited states.
The SI-EE-ADC results exhibit a different dependence on
perturbation order: ADC(2)-X provides the closest agreement
with the experimental $^2\Delta_{3/2u}$ energy, whereas
ADC(3) underestimates this splitting but avoids the
pronounced overestimation of higher excitation energies
observed with SI-EA-ADC(3).
These contrasting trends highlight the sensitivity of
the predicted spin--orbit energy levels to the underlying
ADC formulation and treatment of electron correlation.

We further assess SI-ADC for the lanthanide complex
CeCl$_6^{3-}$, which calculated and experimental
energy-level structures are compared in \cref{fig:CeCl6}.
As for UO$_2^+$, the choice of ADC perturbation order
has a substantially greater impact on the predicted
energy levels than the choice between BP and XDKH1.
Both SI-EA-ADC(2) and SI-EA-ADC(2)-X reproduce the
overall experimental energy-level structure, with
ADC(2)-X providing consistently better agreement
with the observed splittings.
In contrast, ADC(3) substantially overestimates
the excitation energies and produces an excessively
large separation between the lower and upper
energy-level manifolds, failing to reproduce the
experimental spectrum.
The similar behavior of BP and XDKH1 indicates that
these large errors originate primarily from the
underlying ADC description of the spin-free
electronic states rather than the choice of
SOC Hamiltonian, as discussed in \cref{sec:results_5}.

Overall, these benchmarks demonstrate that second-order
SI-ADC can provide a reasonable description of
spin--orbit energy levels even in heavy $f$-block
compounds. Nevertheless, the strong relativistic effects
in these systems challenge the validity of the perturbative
state-interaction treatment, and its apparent accuracy
may partly reflect fortuitous error cancellation.
Therefore, SI-ADC should be applied with caution to
heavy-element compounds, for which two- or four-component
relativistic approaches that incorporate SOC directly
into the electronic structure treatment are generally
more appropriate.

\section{Conclusion}
\label{sec:ctonclusion}

In this work, we have systematically assessed the accuracy and applicability of the state--interaction algebraic diagrammatic construction (SI-ADC) approach for describing spin--orbit coupling (SOC) in molecular systems. Our benchmarks demonstrate that SI-ADC can achieve accuracy comparable to more computationally demanding approaches, including variational four-component ADC and methods that incorporate SOC directly into the ADC effective Hamiltonian. In particular, second-order SI-ADC provides accurate zero-field splittings (ZFS) across a broad range of main-group and transition-metal systems. Although the Breit--Pauli (BP) Hamiltonian occasionally yields more accurate ZFS values than the relativistically more complete XDKH1 Hamiltonian, this behavior is likely attributable to fortuitous error cancellation rather than a systematic advantage of BP.

We have further demonstrated the versatility of SI-ADC by extending its applications to core-level ionization energies and molecular magnetic properties. While absolute core-ionization energies become less accurate for heavier elements, SI-ADC provides reliable SOC-induced splittings of core-ionized states. Moreover, second-order SI-ADC accurately reproduces experimental electronic $g$-tensors for a diverse set of molecules and transition-metal complexes, with XDKH1 offering improved accuracy for systems containing heavy $p$-shell elements. In contrast, SI-ADC(3) exhibits substantial errors for several transition-metal complexes, which can be traced primarily to deficiencies in the underlying spin-free ADC(3) description of electron-attached and ionized states in moderately or strongly correlated systems. Our benchmarks of $f$-block compounds further highlight the limitations of the perturbative state-interaction treatment in the strong-SOC regime, where two- or four-component relativistic approaches are generally more appropriate.

Overall, our results establish SI-ADC as an efficient and versatile framework for incorporating SOC into correlated electronic structure calculations, providing access to a broad range of spectroscopic and magnetic properties at a moderate computational cost. The limitations identified in this work also suggest several promising directions for future development. In particular, extending the state-interaction framework to multireference ADC (MR-ADC) would enable a more balanced description of strongly correlated systems, orbitally degenerate states, and
complex open-shell electronic structures.\cite{sokolov:2018p204113,chatterjee:2019p5908,chatterjee:2020p6343,mazin:2021p6152} 
Complementary developments of spin-adapted single-reference ADC formulations could mitigate spin contamination and improve the reliability of SOC matrix elements for open-shell systems. Together with extensions to a wider range of spectroscopic observables, including SOC effects in excited-state and core-level spectroscopies, these advances would further expand the scope of SI-ADC and establish it as a general framework for investigating relativistic effects in molecular electronic structure and spectroscopy.

\suppinfo
Molecular geometries, computational details, and numerical benchmark results for zero-field splittings, core-ionization energies, electronic $g$-tensors, and spin--orbit-coupled energy levels.

\section*{Data and Software Availability}
The data underlying this study are available in the article and the Supporting Information. 
The codes implementing the methods used in this work are openly available in the \textsc{PySCF} (\url{https://github.com/pyscf/pyscf}) and \textsc{Prism} (\url{https://github.com/sokolov-group/prism}) GitHub repositories.

\section*{Acknowledgements}
This material is based upon work supported by the U.S. Department of Energy, Office of Science, Office of Basic Energy Sciences, Chemical Sciences, Geosciences, and Biosciences Division, Atomic, Molecular, and Optical Sciences Program, under Award Number DE-SC0026341.
Computations were performed at the Ohio Supercomputer Center under Project No.\@ PAS1963.
\cite{osc1987}


\providecommand{\latin}[1]{#1}
\makeatletter
\providecommand{\doi}
  {\begingroup\let\do\@makeother\dospecials
  \catcode`\{=1 \catcode`\}=2 \doi@aux}
\providecommand{\doi@aux}[1]{\endgroup\texttt{#1}}
\makeatother
\providecommand*\mcitethebibliography{\thebibliography}
\csname @ifundefined\endcsname{endmcitethebibliography}
  {\let\endmcitethebibliography\endthebibliography}{}

\end{document}